\documentclass[a4paper]{article}

\usepackage[pages=all, color=black, position={current page.south}, placement=bottom, scale=1, opacity=1, vshift=5mm, contents={}]{background}

\usepackage[margin=1in]{geometry} 

\usepackage{amsmath}
\usepackage{amsthm}
\usepackage{amssymb}
\usepackage{caption}
\usepackage{subcaption}
\usepackage{tabularx}
\usepackage{makecell}
\usepackage{ctable}
\usepackage{multirow}
\usepackage{stmaryrd}
\usepackage{textgreek}
\usepackage{graphicx, color}

\usepackage[utf8]{inputenc}
\usepackage{hyperref}
\hypersetup{
	pdfauthor={Victor Grand, Baptiste Flipon, Alexis Gaillac, Marc Bernacki},
	pdftitle={Characterization and modeling of the influence of initial microstructure on recrystallization of zircaloy-4 during hot forming},
	pdfsubject={Recrystallization of zircaloy-4},
	pdfkeywords={Zircaloy-4, Simulation, Characterization, Recrystallization},
	pdfproducer={LaTeX},
	pdfcreator={pdflatex}
}

\title{Characterization and modeling of the influence of initial microstructure on recrystallization of zircaloy-4 during hot forming}
\author{Victor Grand$^{1,2}$ \and Baptiste Flipon$^1$ \and Alexis Gaillac$^2$  \and Marc Bernacki$^1$ }

\date{
	$^1$Mines ParisTech, PSL - Research University, CNRS, CEMEF, UMR 7635 \\
	$^2$Framatome - Components Research Center\\[2ex]%
}

\begin{document}
	\maketitle
	
	\begin{abstract}
The present article proposes a detailed study of recrystallization of zircaloy-4 under hot forming conditions by means of experimental and numerical tools. Thermomechanical tests and characterization campaigns that have been necessary for this work are described. Then, the different microstructure evolution mechanisms are characterized, from the simplest one to the most complex. Grain growth kinetics is quantified and the influence of second phase particle population is analyzed. Then, a complete study of dynamic and post-dynamic recrystallization is provided. The occurrence of a continuous mechanism is confirmed and the influence of thermomechanical conditions upon recrystallization is assessed. Later, the numerical framework used to simulate grain growth, continuous and post-dynamic recrystallization is presented. After having successfully reproduced the grain coarsening kinetics with and without second phase particles, the model is used to describe continuous dynamic recrystallization and post-dynamic recrystallization from an initial equiaxed and fully recrystallized microstructure. The agreement between experimental and numerical results is assessed in details. Finally, post-dynamic recrystallization is simulated, starting from two deformed microstructures characterized by electron back-scattered diffraction technique and immersed into simulations. This allows to discuss and reproduce the influence of initial microstructure.		
	\end{abstract}

\newpage

\tableofcontents

\newpage

\section{Introduction}\label{sec:Introduction}

With the global technology advances, requirements for metallic materials have also known a significant increase. Thus, a need for a better understanding of metallurgy, microstructure evolutions and the link between microstructure characteristics and macroscopic properties is required. This is especially true for critical applications such as a aeronautics or nuclear industry. Focusing on nuclear industry, the development of new reactors and the general regulation reinforcement oblige to fully understand and master every single processing step to ensure that each final product meets the high requirements. The fabrication of zirconium alloy cladding fuel assembly follows this principle. Thus, for several decades now, Framatome has been leading research campaigns to improve knowledge about zirconium alloy properties \cite{Bossis2001,ToffolonMasclet2008}, forming processes and microstructure evolutions \cite{Chauvy2006, Vanderesse2008_PhD, Ben_Ammar2012_PhD}.

Manufacturing route of tubes and grids consists in a succession of hot and cold forming steps and heat treatments. Each of these stages is crucial since it impacts both the following process and the final part. It influences the piece geometry as well as some macroscopic properties such as ductility, mechanical strength or even corrosion resistance. These characteristics being intimately related to microstructural features such as grain size distribution or second phase particle (SPP) properties. Consequently, understanding and being able to predict how microstructure evolves during hot forming and heat treatments, especially through recrystallization (ReX), is of first interest. The particular case of hot extrusion used to produce cladding tubes illustrates how this is critical. Indeed, a low recrystallized fraction after hot forming is susceptible to induce crack formation during the subsequent steps of cold pilgering \cite{Gaillac2018, Gaillac2011}.

Recrystallization of zircaloy-4 (Zy-4) has been studied by the way of experimental characterization campaigns in several studies in the past. Chauvy et al. \cite{Chauvy2006} have put in light that under common industrial hot deformation conditions, Zy-4 exhibits continuous dynamic recrystallization (CDRX). This mechanism is characterized by the formation of subgrains and their progressive transformation into new recrystallized grains. The new recrystallized grains are generally characterized by a low dislocation density and an equiaxed shape \cite{Rollett2017}. During CDRX, they appear rather homogeneously throughout the whole microstructure \cite{Huang2016}. Additionally, Chauvy et al. have analyzed quantitatively the formation of low angle grain boundaries (LAGB, i.e. grain boundaries (GB) that have a disorientation lower than a given threshold value, generally taken equal to $15 ^{\circ}$) and the influence of thermomechanical conditions on their formation \cite{Chauvy2006}. Logé et al. \cite{Loge2000} have investigated in details the influence of the lamellae thickness and of the precipitate distribution upon texture evolution during hot deformation and its relation with recrystallization. They exhibit how such initial microstructural features may affect subsequent evolutions during processing.

As it has been shown by experimental studies, recrystallization of Zy-4 is influenced by a high number of parameters \cite{Chauvy2006, Vanderesse2008_PhD, Loge2000}. Therefore, numerical modeling appears as a suitable tool to assess separately the influence of one or the other condition. Dunlop et al., in a series of two articles, have developed models to predict the plastic deformation and the recrystallization of zircaloy-4 \cite{Dunlop2007_PlaticDef, Dunlop2007_Rx}. They propose a mean field model, i.e. a model based on variables describing the average characteristics of the considered microstructure, for both recovery and recrystallization, which predicts successfully the recrystallization kinetics. Gaudout et al. \cite{Gaudout2009} have adapted the Gourdet-Montheillet model \cite{Gourdet2003} to simulate the recrystallization of zirconium alloys during and after hot deformation. This last study constitutes at the moment the last and more detailed attempt to model CDRX of Zy-4.

Until now and to the authors knowledge, no study has been led to assess in details the influence of initial microstructure on recrystallization of Zy-4. All modeling efforts have been made to produce a consistent model of continuous recrystallization. The present article intends to take advantage of the recent advances made in full-field simulations in order to improve our understanding of CDRX of Zy-4. It summarizes a complete and extensive study of Zy-4 recrystallization during and after hot forming in conditions representative of industrial processes \cite{Gaillac2018}. First, it describes how microstructure evolutions have been characterized and what are the main takeaways about recrystallization mechanisms. It discusses to what extent thermomechanical conditions and initial microstructure impact recrystallization. Secondly, this paper presents the full-field simulation framework used to model grain growth (GG), CDRX and post-dymamic recrystallization (PDRX). Then, the results from the main simulation cases are presented and a discussion upon the influence of several microstructure characteristics is proposed.

\section{Materials and methods}\label{sec:MaterialsAndMethods}

\subsection{Materials}\label{subsec:Materials}

Two different materials corresponding to three characteristic microstructures are used in the experimental campaigns presented in the following sections:
\begin{itemize}
\item commercial zircaloy-4, with either equiaxed grains or a lamellar microstructure. Samples with lamellar microstructure have been manufactured for this study and exhibit mainly basket-weaved lamellae.
\item A zirconium alloy with a chemical composition as close as possible to the specification of Zy-4, except for Iron and Chromium that are kept as low as possible. Samples made out of this alloy are manufactured to obtain an alloy with the same solid solution than Zy-4 but without SPP.
\end{itemize}

\begin{figure}[h!]
	\centering
 	\includegraphics[width=0.6\textwidth]{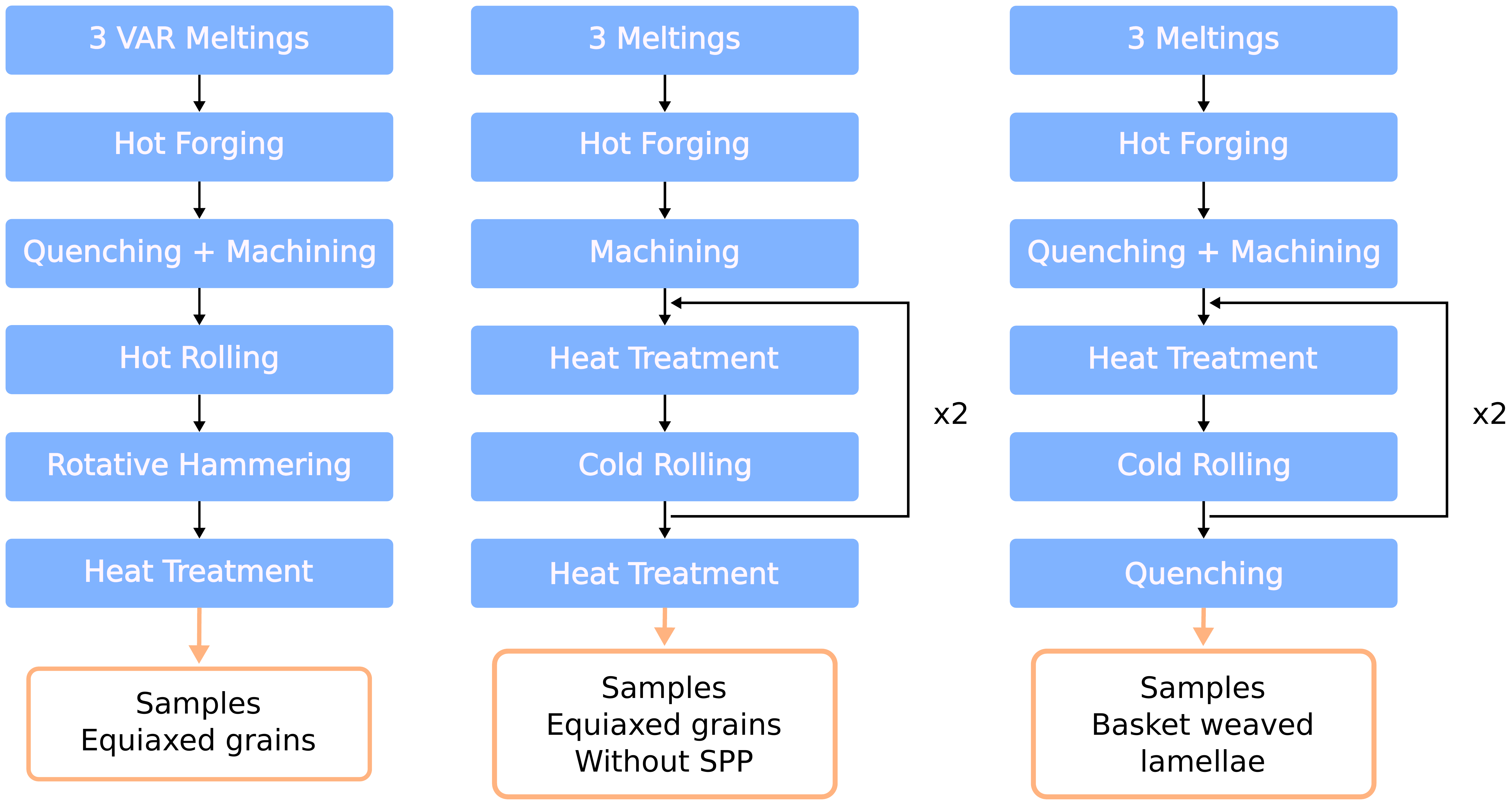}
  \caption{Presentation of the manufacturing routes of each material.}\label{fig:ManufacturingRoutes}
\end{figure}

For the sake of clarity, those three different initial states will be respectively called: equiaxed (eq.), basket-weaved (b-w.) and without SPP (no SPP.).
The manufacturing routes of each of those part are described in figure \ref{fig:ManufacturingRoutes}.

Due to the fact each of those three different initial states are not obtained by the exact same processing route, it is not possible to manufacture cylindrical samples for thermomechanical testing with the exact same geometry. Nevertheless, the three geometries share the same cylindrical shape with a ratio height over radius being equal to 1.4.

\subsection{Thermomechanical testing}\label{subsec:TTh}

Both heat treatments and hot compression tests are made within this study. Heat treatments are made either under vacuum or a protective atmosphere of argon to limit surface oxidation.
Hot compression experiments are made using a hydraulic mechanical testing machine ($\textit{MTS}$  $\textit{Landmark}$ $\textit{370-25}$). A radiant furnace is used to heat samples and dies at the temperature setpoint. The set-up allows the furnace to move along compression direction in order to put sample in place before experiment and to proceed quickly to the water quench at the end of the experiment. Dies and samples are coated using a spray of boron nitride. The whole test is filmed to determine the quenching time for each experiment. Detailed tables presenting the conditions of thermomechanical tests are available in table \ref{tab:TestMatrix}.

\begin{table}[h]
    \begin{subtable}{0.95\textwidth}
        \centering
        \centering
        \begin{tabular}{l c c c c c c c c c}\toprule
        Strain & \multicolumn{3}{c}{$\mathbf{0.01 ~ s^{-1}}$} & \multicolumn{3}{c}{$\mathbf{0.1 ~ s^{-1}}$} & \multicolumn{3}{c}{$\mathbf{1.0 ~ s^{-1}}$} \\
        \midrule
        $\mathbf{450 ^{\circ}C}$ & 0.65 & 1.0 & 1.35 & 0.65 & 1.0 & 1.35 & 0.65 & 1.0 & 1.35\\
        $\mathbf{550 ^{\circ}C}$ & 0.65 & 1.0 & 1.35 & 0.65 & 1.0 & 1.35 & 0.65 & 1.0 & 1.35\\
        $\mathbf{650 ^{\circ}C}$ & 0.65 & 1.0 & 1.35 & 0.65 & 1.0 & 1.35 & 0.65 & 1.0 & 1.35\\
        \bottomrule
        \end{tabular}
        \caption{\label{tab:TestMatrixHCR} Hot compression with equiaxed microstructures.}
    \end{subtable}
    \hfill
    \begin{subtable}{0.95\textwidth}
        \centering
        \begin{tabular}{l c c c c c c c c c}\toprule
        Strain & \multicolumn{3}{c}{$\mathbf{0.01 ~ s^{-1}}$} & \multicolumn{3}{c}{$\mathbf{0.1 ~ s^{-1}}$} & \multicolumn{3}{c}{$\mathbf{1.0 ~ s^{-1}}$} \\
        \midrule
        $\mathbf{450 ^{\circ}C}$ & 0.65 & 1.0 & 1.35 & \textcolor{gray}{\textbf{-}} & \textcolor{gray}{\textbf{-}} & 1.35 & \textcolor{gray}{\textbf{-}} & \textcolor{gray}{\textbf{-}} & \textcolor{gray}{\textbf{-}} \\
        $\mathbf{550 ^{\circ}C}$ & \textcolor{gray}{\textbf{-}} & \textcolor{gray}{\textbf{-}} & 1.35 & 0.65 & 1.0 & 1.35 & \textcolor{gray}{\textbf{-}} & \textcolor{gray}{\textbf{-}} & 1.35\\
        $\mathbf{650 ^{\circ}C}$ & \textcolor{gray}{\textbf{-}} & \textcolor{gray}{\textbf{-}} & \textcolor{gray}{\textbf{-}} & \textcolor{gray}{\textbf{-}} & \textcolor{gray}{\textbf{-}} & 1.35 & 0.65 & 1.0 & 1.35\\
        \bottomrule
        \end{tabular}
        \caption{\label{tab:TestMatrixHCVandHCP} Hot compression with lamellar microstructures.}
     \end{subtable}
     \begin{subtable}{0.95\textwidth}
     \centering
    \begin{tabular}{l l c c c c c c c c c c c c}\toprule
    \multicolumn{2}{c}{Holding time (s)} & \multicolumn{6}{c}{$\mathbf{0.1 ~ s^{-1}}$} & \multicolumn{6}{c}{$\mathbf{1.0 ~ s^{-1}}$} \\
    \midrule
    \multirow{2}{*}{$\mathbf{650 ^{\circ}C}$ } & \textbf{0.65} & \textcolor{gray}{\textbf{-}} & \textcolor{gray}{\textbf{-}} & \textcolor{gray}{\textbf{-}} & \textcolor{gray}{\textbf{-}} & \textcolor{gray}{\textbf{-}} & \textcolor{gray}{\textbf{-}} & 7 & 12 & 25 & 50 & 100 & 200 \\
    & \textbf{1.35} & 7 & 12 & 25 & 50 & 100 & 200 & 7 & 12 & 25 & 50 & 100 & 200 \\
    \bottomrule
    \end{tabular}
    \caption{\label{tab:TestMatrixtHCHR} Hot compression and holding at temperature with equiaxed microstructures.}
     \end{subtable}
     \begin{subtable}{0.95\textwidth}
     \centering
    \begin{tabular}{l l c c c c c c c c }\toprule
    \multicolumn{2}{c}{Holding time (s)} & \multicolumn{4}{c}{$\mathbf{0.1 ~ s^{-1}}$} & \multicolumn{4}{c}{$\mathbf{1.0 ~ s^{-1}}$} \\
    \midrule
    \multirow{2}{*}{$\mathbf{650 ^{\circ}C}$ } & \textbf{0.65} & \textcolor{gray}{\textbf{-}} & \textcolor{gray}{\textbf{-}} & \textcolor{gray}{\textbf{-}} & \textcolor{gray}{\textbf{-}} & 12 & 25 & 50 & 100 \\
    & \textbf{1.35} &  12 & 25 & 50 & 100 & 12 & 25 & 50 & 100\\
    \bottomrule
    \end{tabular}
    \caption{\label{tab:TestMatrixHCHVandHCHP} Hot compression and holding at temperature with lamellar microstructures.}
     \end{subtable}
     \caption{Description of the thermomechanical conditions of the hot compression campaigns.}
     \label{tab:TestMatrix}
\end{table}

\subsection{Characterization techniques}\label{subsec:Characterization}

Two techniques are used to characterize microstructure evolutions: optical microscopy under polarized light with image analysis \cite{Flipon2021} (OM-PL) and EBSD. The first technique is used for characterizing fully recrystallized microstructures such as the ones observed under GG condition. EBSD, on the contrary, is required for studying strain hardened microstructures. EBSD analysis are performed using a \textit{Carl Zeiss SUPRA 40} field emission gun scanning electron microscope (FEG-SEM) equipped with a \textit{Bruker Quantax} system (EBSD \textit{e$^-$Flash$^{\textsc{hr}}$}). Acceleration voltage is set equal to $20 ~ kV$ and step size to $0.1 ~ \mu m$. The step size is kept constant for all samples and is set sufficiently low to be able to capture local orientation heterogeneities such as subgrains that are in general few hundred nanometers wide. Each observation zone dimensions are: $130 ~ \mu m$ in width and $85 ~ \mu  m$ in height. This area contains between $500$ and $3000$ grains which is enough to ensure minimum representativeness.

Samples observed using OM-PL technique are prepared according to the protocol described in ref. \cite{Flipon2021}. Samples observed by EBSD are polished using first grinding papers and a polishing solution that contains diamond particles with a size of $15 ~ nm$. Then, they are electropolished using a solution of $90 \%$ methanol and $10 \%$ perchloric acid. Voltage is set equal to $25 ~ V$, polishing is performed during $27 ~ s$ at a temperature between 5 and 10$^{\circ}C$.

EBSD data are post-processed using MTEX \cite{Bachmann2011}. LAGB are considered to have a disorientation within $\left[ 2 ^{\circ} ~ ; ~ 15 ^{\circ} \right]$. LAGB that are enclosing subgrains are designated as closed LAGB. All data have been filtered using a half-quadratic filter to reduce noise \cite{Hielscher2019}. Geometrically necessary dislocation (GND) density is estimated following the method developed by Pantleon \cite{Pantleon2008}.
Finally, a criterion based on average GND density per grain is used for discrimination of recrystallized grains : $\rho_{\textsc{gnd}} \leq 1.0 \times 10^{14} m^{-2}$. This choice is motivated to ensure thorough comparison between experimental and numerical results. Indeed, the other common variables used for identification of recrystallized grains, such as grain orientation spread (GOS) or grain average kernel average misorientation (GAKAM), are not computed in the numerical framework.

\section{Characterization of microstructure evolutions}\label{sec:Charac}

\subsection{Grain growth}\label{subsec:CharacGG}

\begin{figure}
    \centering
    \begin{subfigure}{0.49\textwidth}
        \centering
        \includegraphics[width=0.95\linewidth]{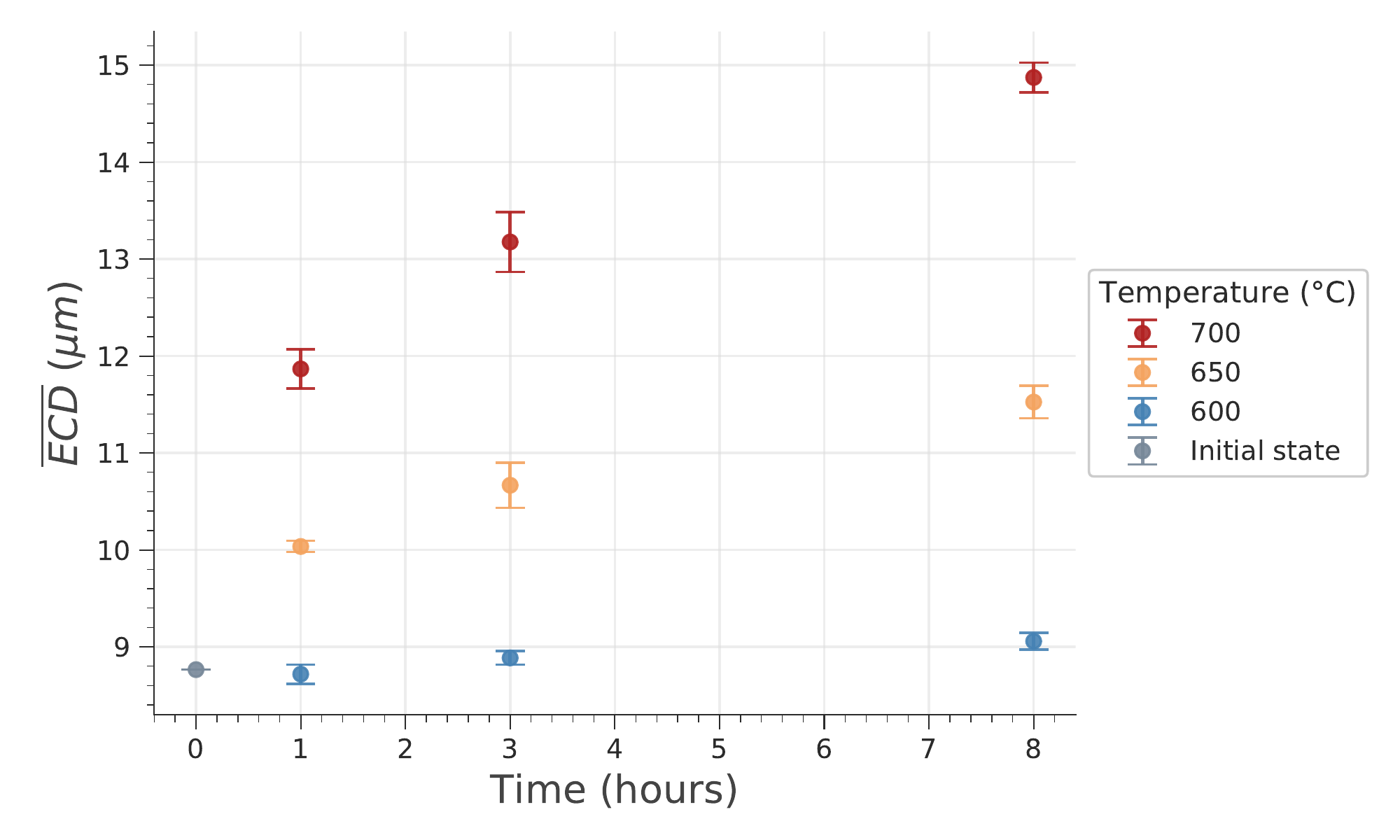}
        \caption{\label{fig:ECDwithTimeNoSPP} Evolution of $\overline{ECD}$ for the alloy without SPP.}
    \end{subfigure}
    \begin{subfigure}{0.49\textwidth}
        \centering
        \includegraphics[width=0.95\linewidth]{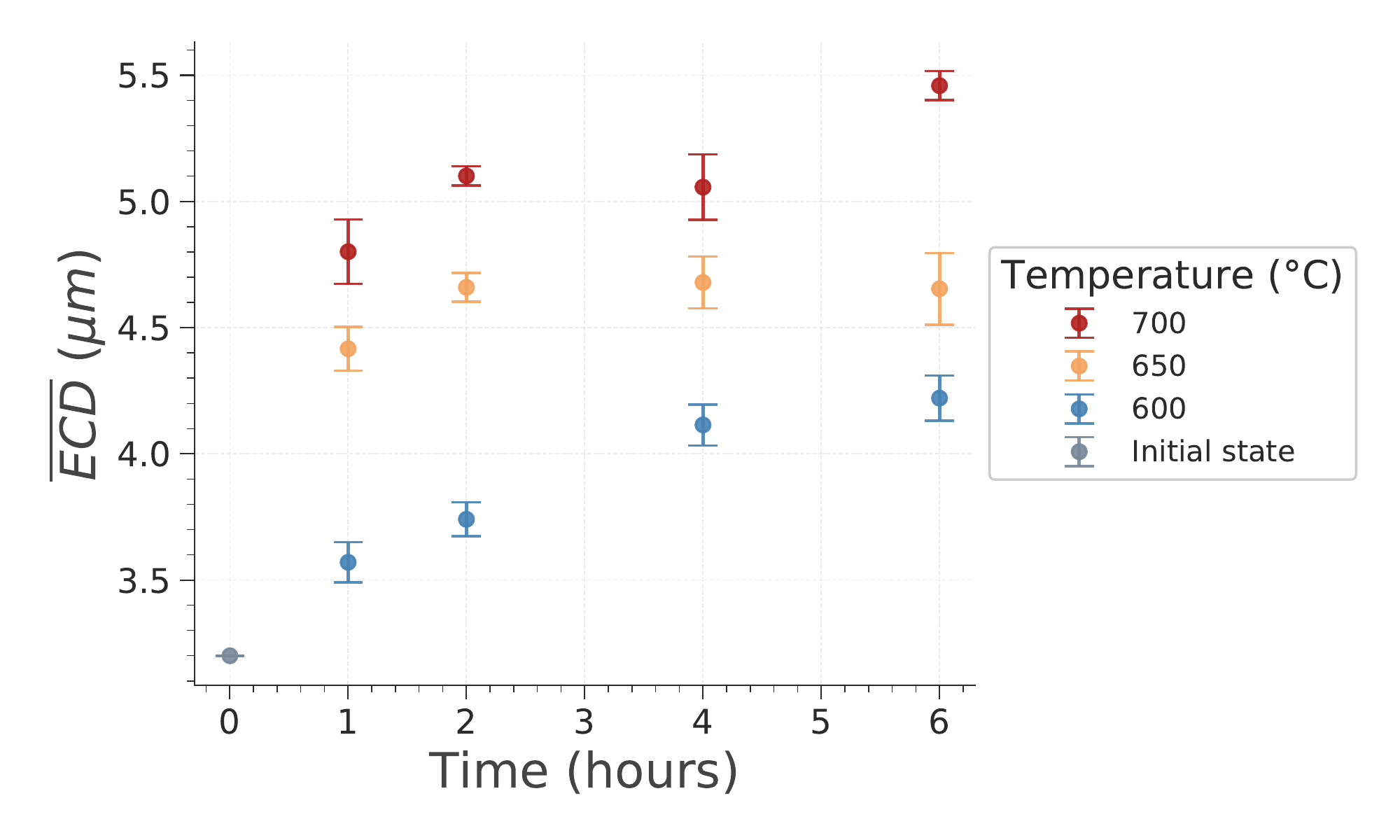}
        \caption{\label{fig:ECDwithTimeWithSPP} Evolution of $\overline{ECD}$ for Zy-4 with SPP.}
    \end{subfigure}
\caption{Evolution of $\overline{ECD}$ as a function of holding time at temperature.}
\label{fig:ExpResultsGG}
\end{figure}

The numerical framework described below (section \ref{sec:NumericalFramework}) requires to characterize grain boundary mobility in the absence of SPP. However, there is no temperature range within $\alpha$-phase stability domain where SPP are soluble \cite{ToffolonMasclet2008}. Consequently, some material have been manufactured with as little as possible SPP fraction. For both eq. and no SPP. initial states, heat treatments are conducted under three temperatures: $600$, $650$ and $700 ^{\circ}C$. For the material without SPP, holding times are $1$, $3$ and $8$ hours whereas for the material with SPP they are $1$, $2$, $4$ and $6$ hours. Figure \ref{fig:ExpResultsGG} presents the arithmetic average of the grain equivalent circle diameter ($\overline{ECD}$) as a function of the heat treatment duration.

\begin{figure}
    \centering
    \begin{subfigure}{0.4\textwidth}
        \centering
        \includegraphics[width=0.8\linewidth]{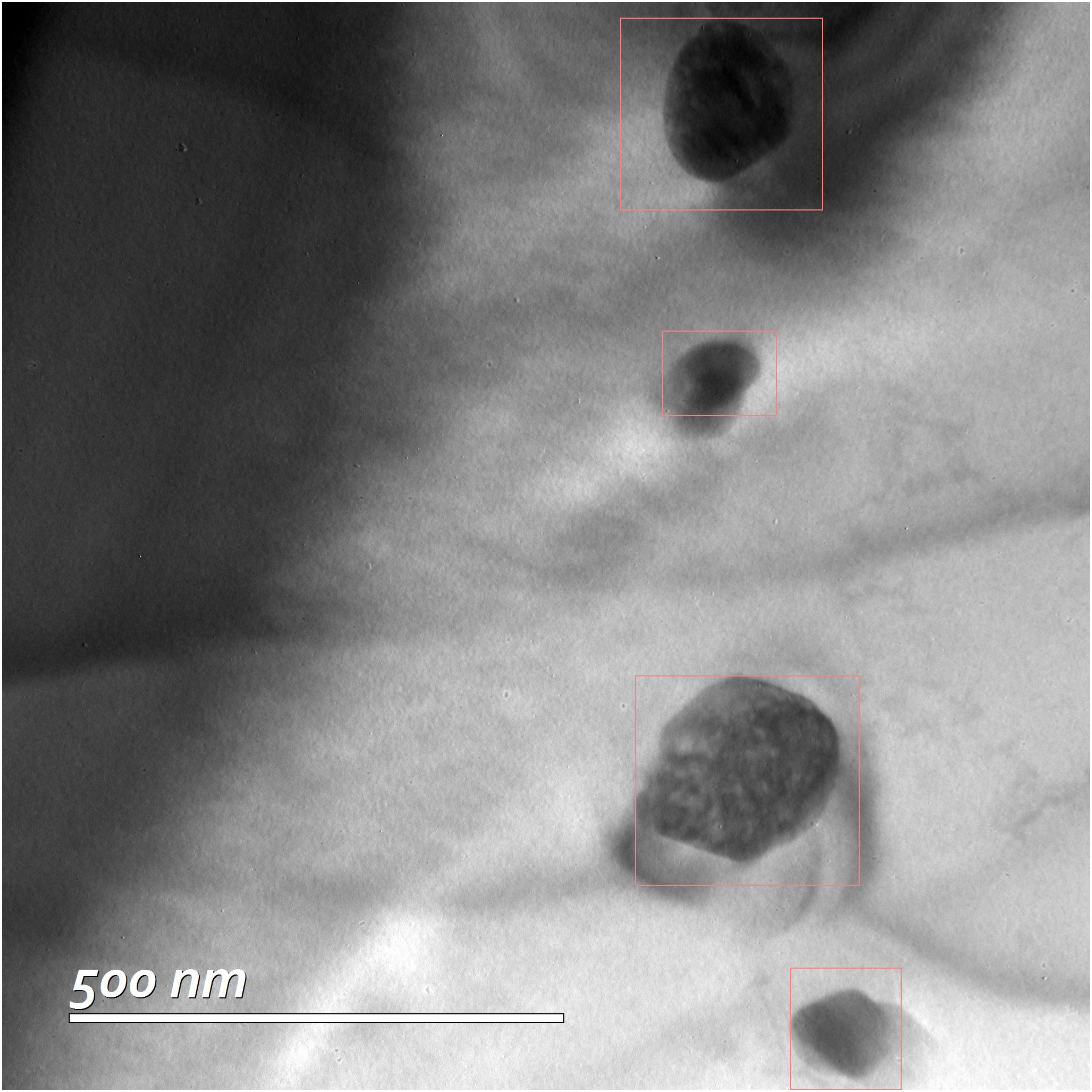}
        \caption{\label{fig:TEMSPPimage} TEM image of SPP.}
    \end{subfigure}
    \begin{subfigure}{0.55\textwidth}
        \centering
        \includegraphics[width=0.95\linewidth]{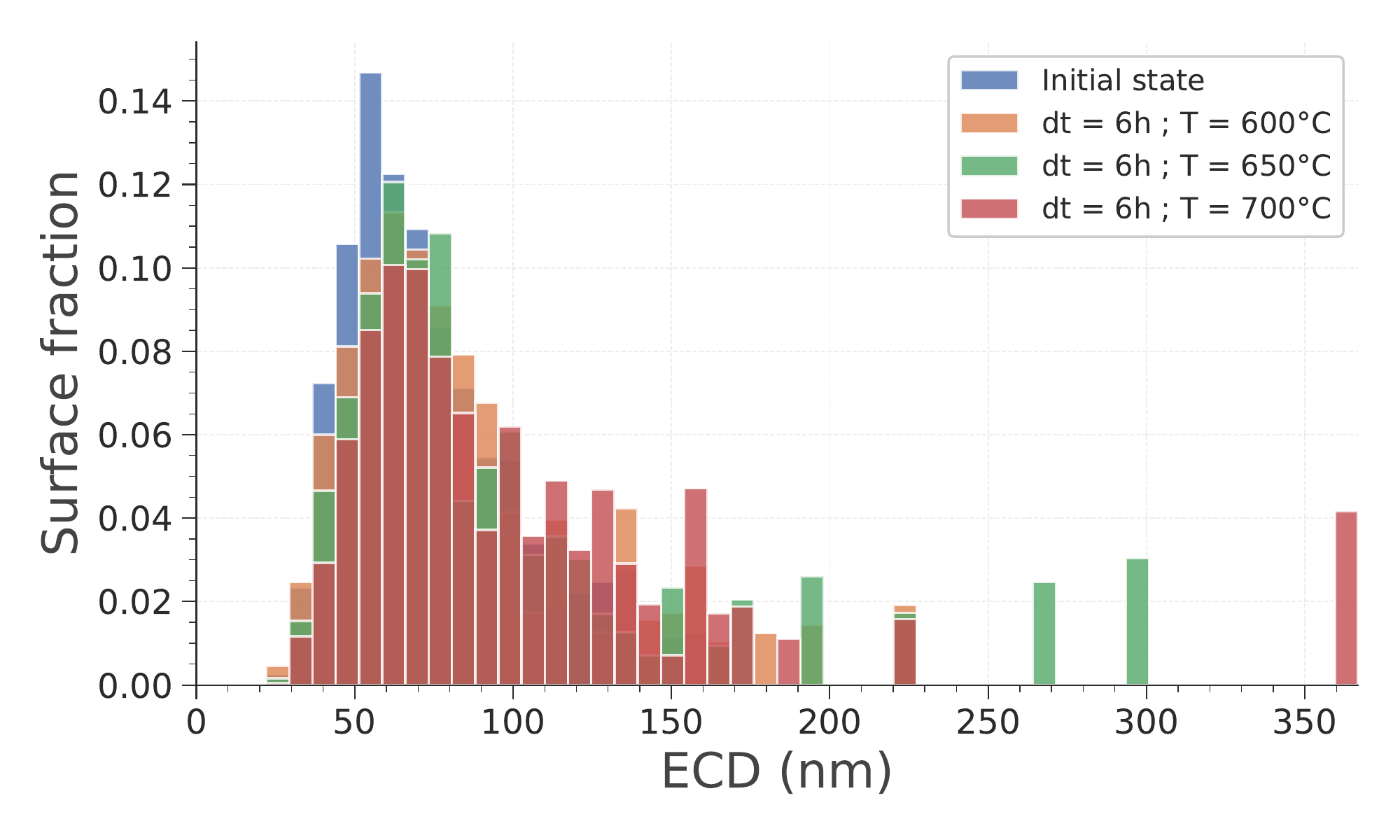}
        \caption{\label{fig:SPPdistribTEM} $ECD$ surface distribution measured by TEM.}
    \end{subfigure}
\caption{Characterization of SPP population by TEM.}
\label{fig:SPPCountingTEM}
\end{figure}

First, it can be seen on figure \ref{fig:ExpResultsGG} that the presence of SPP seems to be responsible for the presence of a plateau for high durations. Those facts are in agreement with the classical theory describing the Smith-Zener pinning effect \cite{Smith1948, Zener1949, Manohar1998}. However, according to the standard Smith-Zener equation, the limit grain size only depends on the SPP population properties that are their volume fraction and their mean equivalent diameter. Consequently, the different plateau values at the three temperatures seem to indicate that the SPP population evolves differently during heat treatments executed at the three considered temperatures. To evaluate this, SPP populations have been characterized by transmission electron microscopy (TEM) for a sample corresponding to the initial state and for three samples corresponding to the final states after 6 hours of heat treatment. Figure \ref{fig:SPPCountingTEM} illustrates how SPP are observed using TEM and the SPP distributions obtained.

One can note that the SPP size distribution does not change that much globally from one heat treatment to the other. However, it is important to keep in mind that some slight changes for low ECD values can greatly influence the number of particles. Those results will be discussed in light of the simulation results in section \ref{subsec:ModelGG}.

\subsection{Recrystallization}\label{subsec:CharacRX}

\subsubsection{Characterization of recrystallization mechanism}\label{subsubsec:RxMechanism}

As said earlier, Zy-4 is supposed to undergo CDRX under the common industrial thermomechanical conditions \cite{Chauvy2006, Gaudout2009}. However, previous studies have been conducted only with Zy-4 presenting a lamellar microstructure at initial state and do not provide enough in details microstructure characterization. This study tackles those two remarks and provides a new insight.

\begin{figure}[h!]
    \centering
     \begin{subfigure}{0.32\textwidth}
        \centering
	    \includegraphics[width=0.98\linewidth]{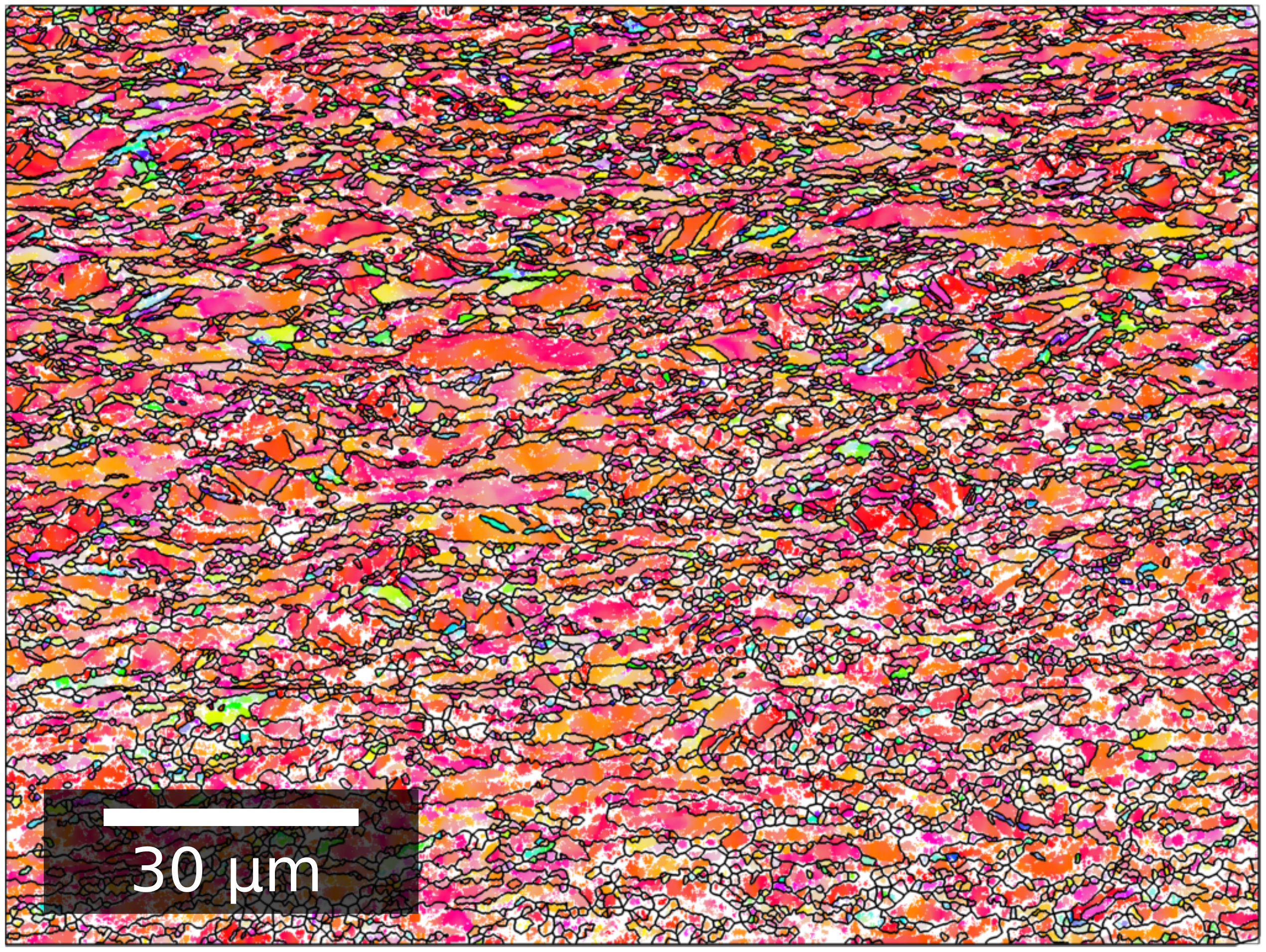}
        \caption{\label{fig:T450_SR1_Strain065_t1s} $T = 450 ^{\circ}$, $\dot{\varepsilon} = 0.01 ~ s^{-1}$, $\varepsilon = 1.35$, $dt = 1 ~ s$.}
    \end{subfigure}
    \vfill
     \begin{subfigure}{0.32\textwidth}
        \centering
	    \includegraphics[width=0.98\linewidth]{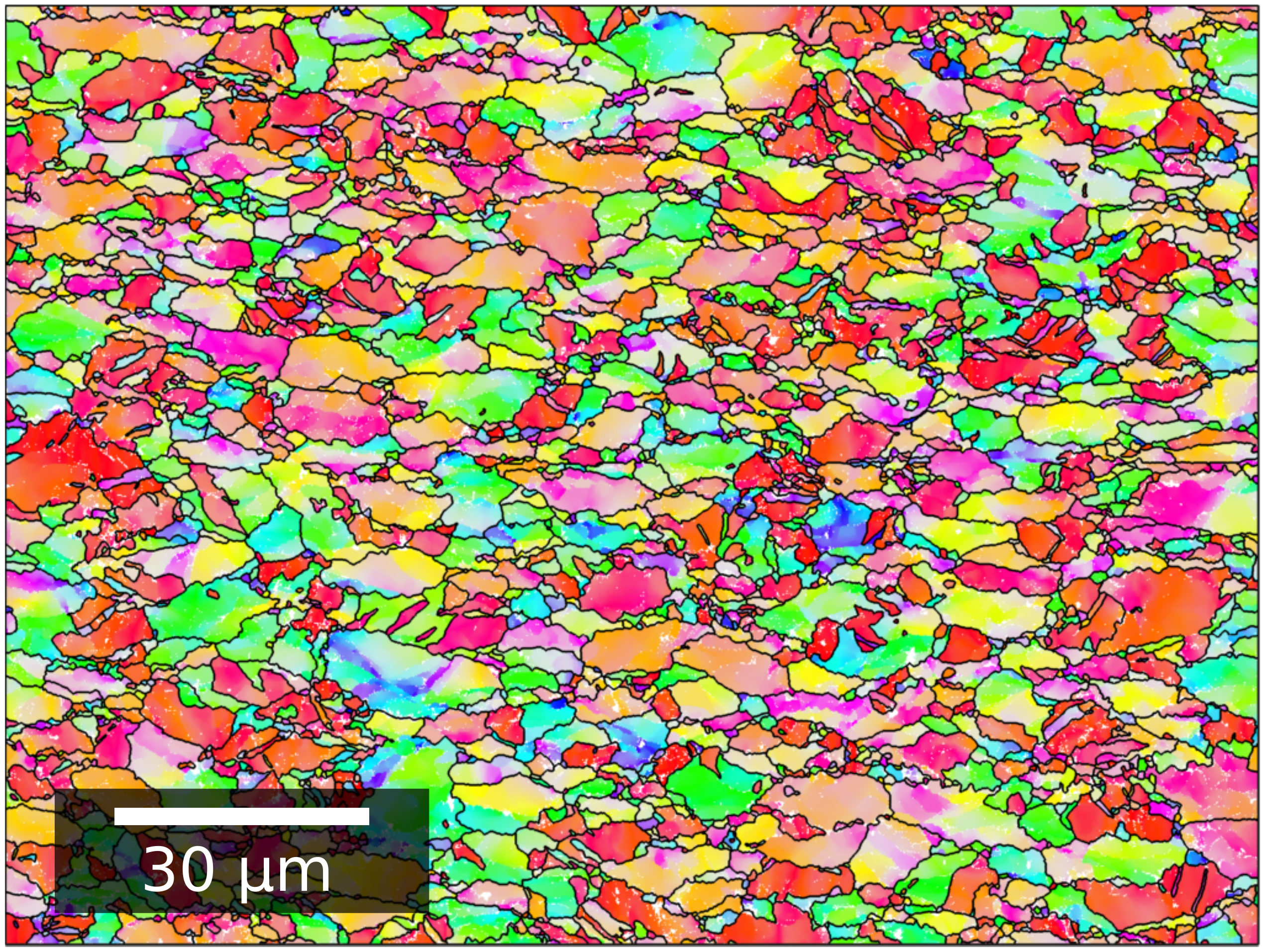}
        \caption{\label{fig:T650_SR1_Strain065_t1s} $T = 650 ^{\circ}$, $\dot{\varepsilon} = 1.0 ~ s^{-1}$, $\varepsilon = 0.65$, $dt = 1 ~ s$.}
    \end{subfigure}
     \begin{subfigure}{0.32\textwidth}
        \centering
	    \includegraphics[width=0.98\linewidth]{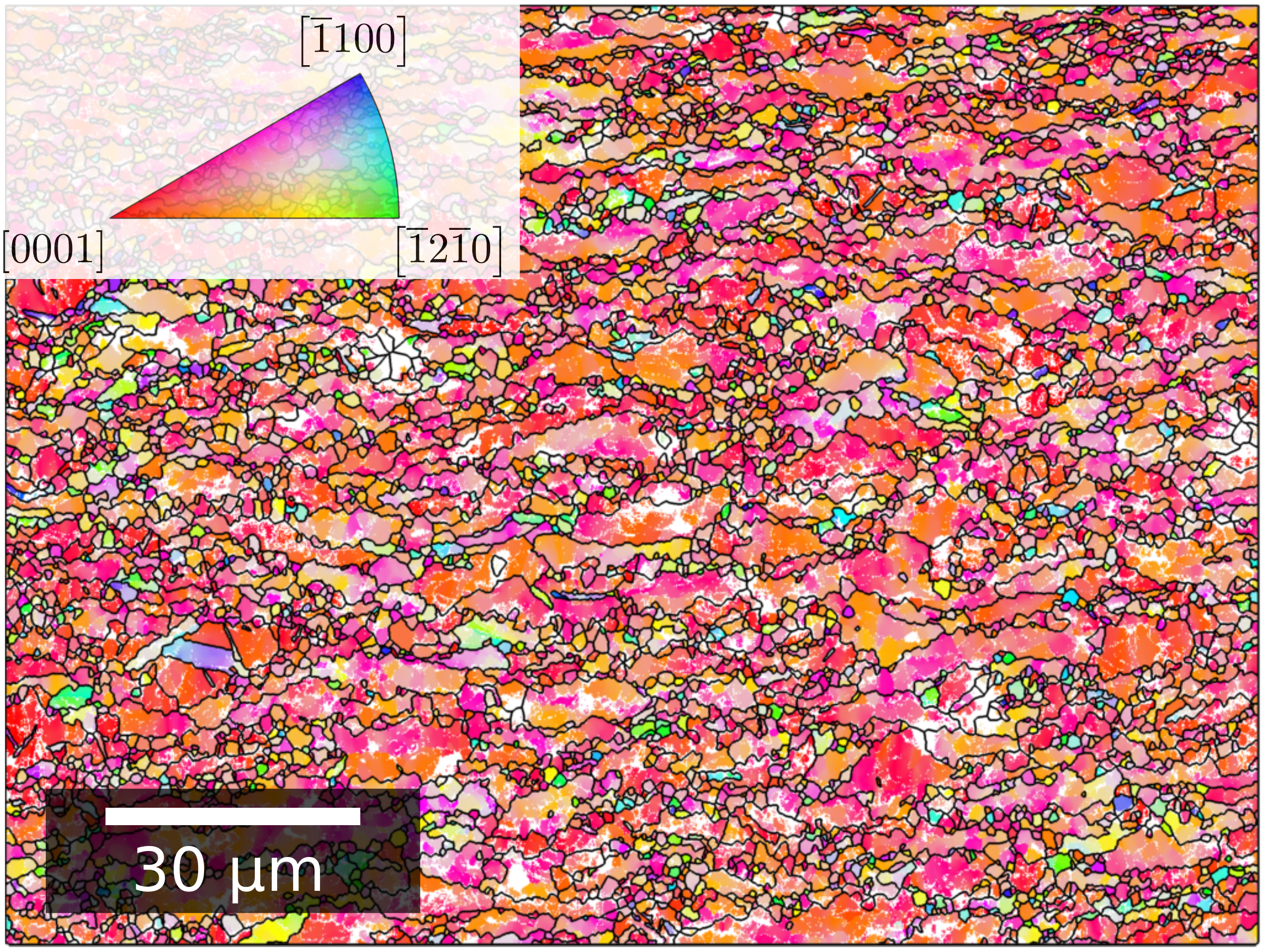}
        \caption{\label{fig:T650_SR1_Strain135_t1s} $T = 650 ^{\circ}$, $\dot{\varepsilon} = 1.0 ~ s^{-1}$, $\varepsilon = 1.35$, $dt = 1 ~ s$.}
    \end{subfigure}
     \begin{subfigure}{0.32\textwidth}
        \centering
	    \includegraphics[width=0.98\linewidth]{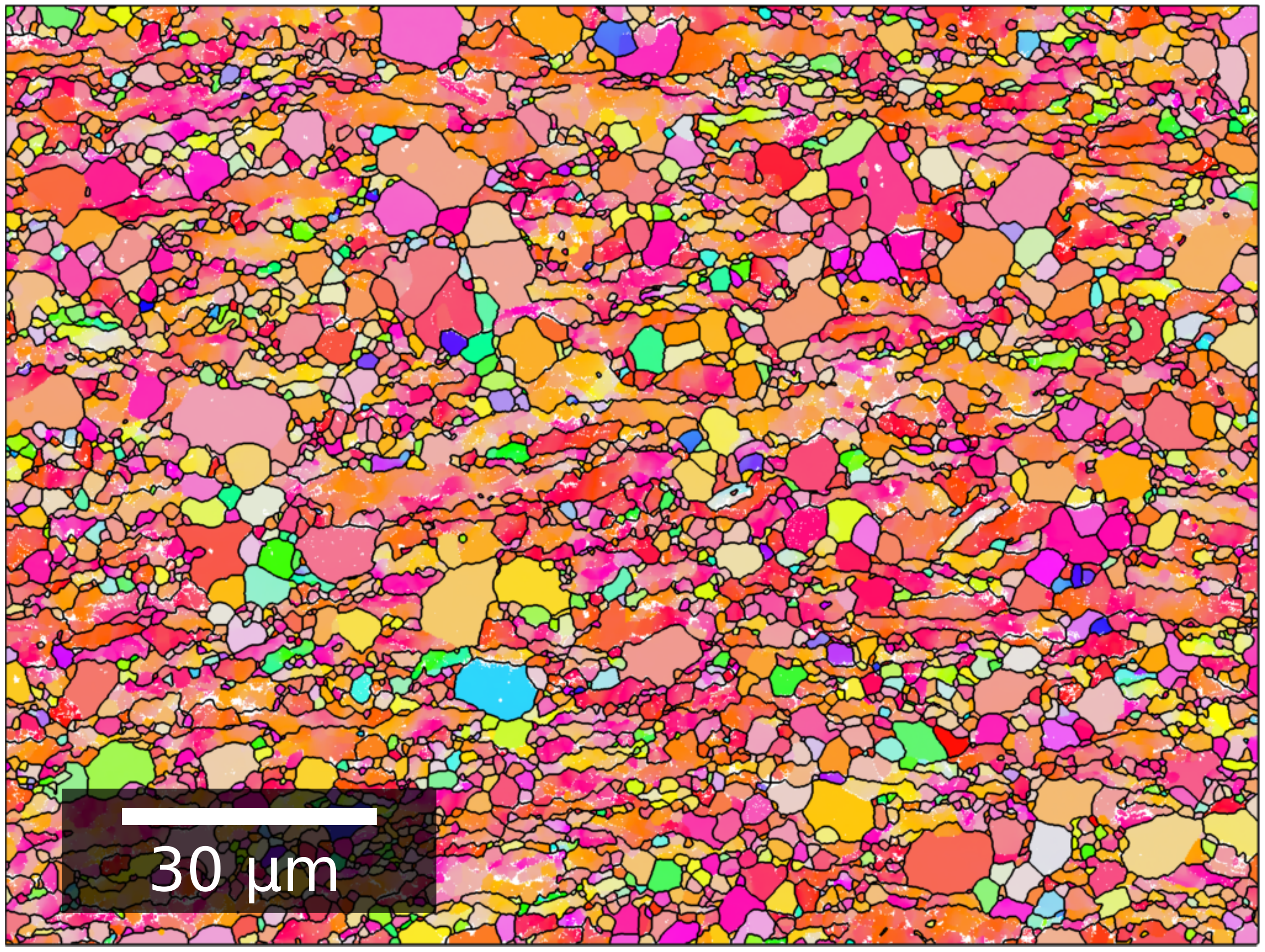}
        \caption{\label{fig:T650_SR1_Strain135_t50s} $T = 650 ^{\circ}$, $\dot{\varepsilon} = 1.0 ~ s^{-1}$, $\varepsilon = 1.35$, $dt = 50 ~ s$.}
    \end{subfigure}
    \vfill
     \begin{subfigure}{0.32\textwidth}
        \centering
	    \includegraphics[width=0.98\linewidth]{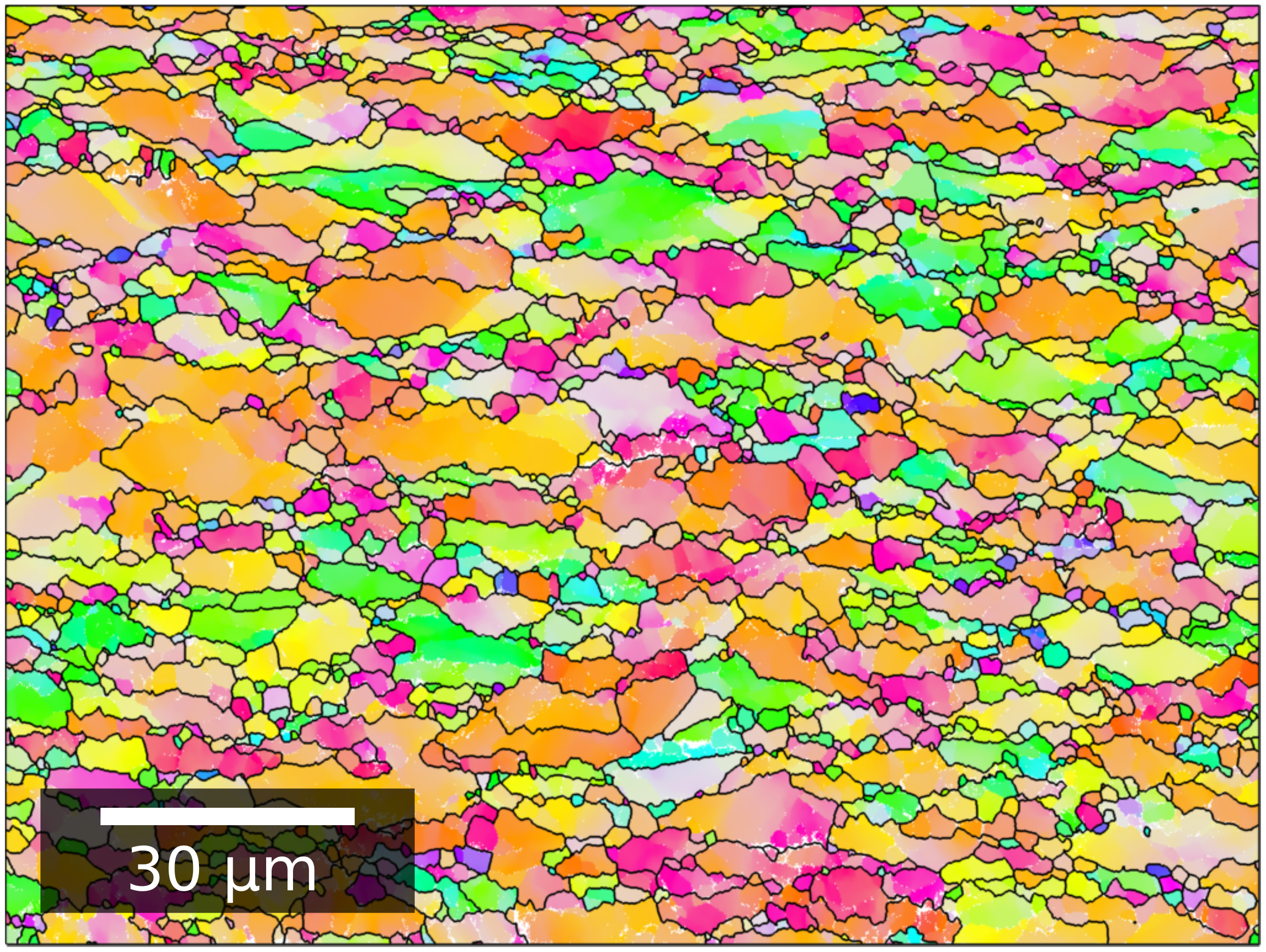}
        \caption{\label{fig:T650_SR001_Strain135_t1s} $T = 650 ^{\circ}$, $\dot{\varepsilon} = 0.01 ~ s^{-1}$, $\varepsilon = 1.35$, $dt = 1 ~ s$.}
    \end{subfigure}
  \caption{Presentation of orientation maps with IPF Y color code for five different microstructures after hot deformation.}\label{fig:DRXmechanisms}
\end{figure}

Figure \ref{fig:DRXmechanisms} presents four orientation maps obtained using EBSD. An inverse pole figure Y (IPF Y) color coding is used (with Y axis being the compression direction). Thermomechanical conditions are detailed for each map. It is interesting to note that when the sample is quenched right after deformation, the microstructure contains large grains that seem to present orientation gradients and/or subgrains. Those grains are elongated along horizontal direction which is perpendicular to the compression direction. The microstructure also includes a large number of small grains dispersed rather homogeneously. The orientation map in figure \ref{fig:T650_SR1_Strain135_t50s} corresponding to the microstructural state after holding at temperature for $50 s$ shows that a few recrystallized grains seem to grow at the expense of the deformed area. This illustrates that even if a large number of subgrains and small grains are formed during hot deformation, only a limited amount fulfill the energetic and/or kinetics criterion to expand. The low growth kinetics of recrystallized grains is consistent with the observations made in previous section that are supposing that Zy-4 has a low grain boundary mobility (see section \ref{subsec:CharacGG}).

\begin{figure}[h!]
    \centering
     \begin{subfigure}{0.49\textwidth}
        \centering
	    \includegraphics[width=0.95\linewidth]{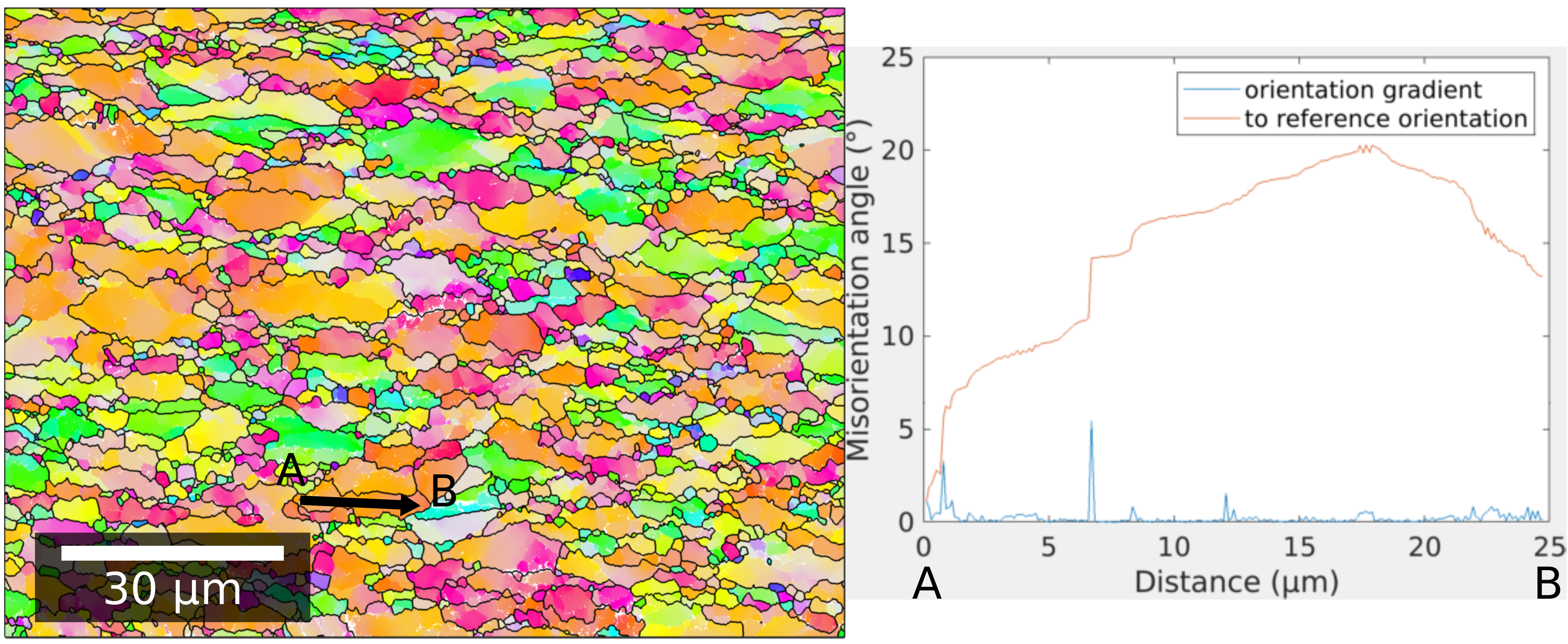}
        \caption{\label{fig:DisorientationProfileGrad} Evolution of disorientation along a given path.}
    \end{subfigure}
    \begin{subfigure}{0.49\textwidth}
        \centering
        \includegraphics[width=0.95\linewidth]{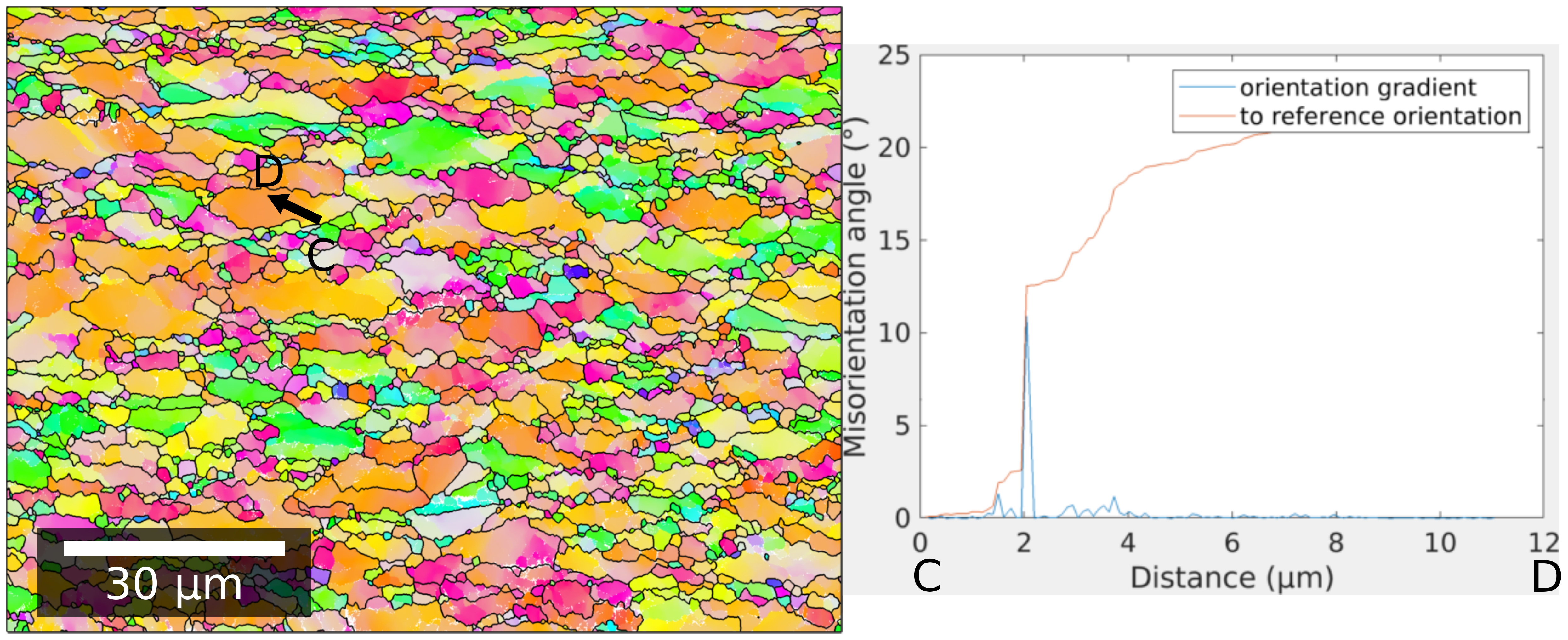}
        \caption{\label{fig:DisorientationProfileMLAGB} Evolution of disorientation along a given path.}
    \end{subfigure}
    \begin{subfigure}{0.49\textwidth}
        \centering
	    \includegraphics[width=0.95\linewidth]{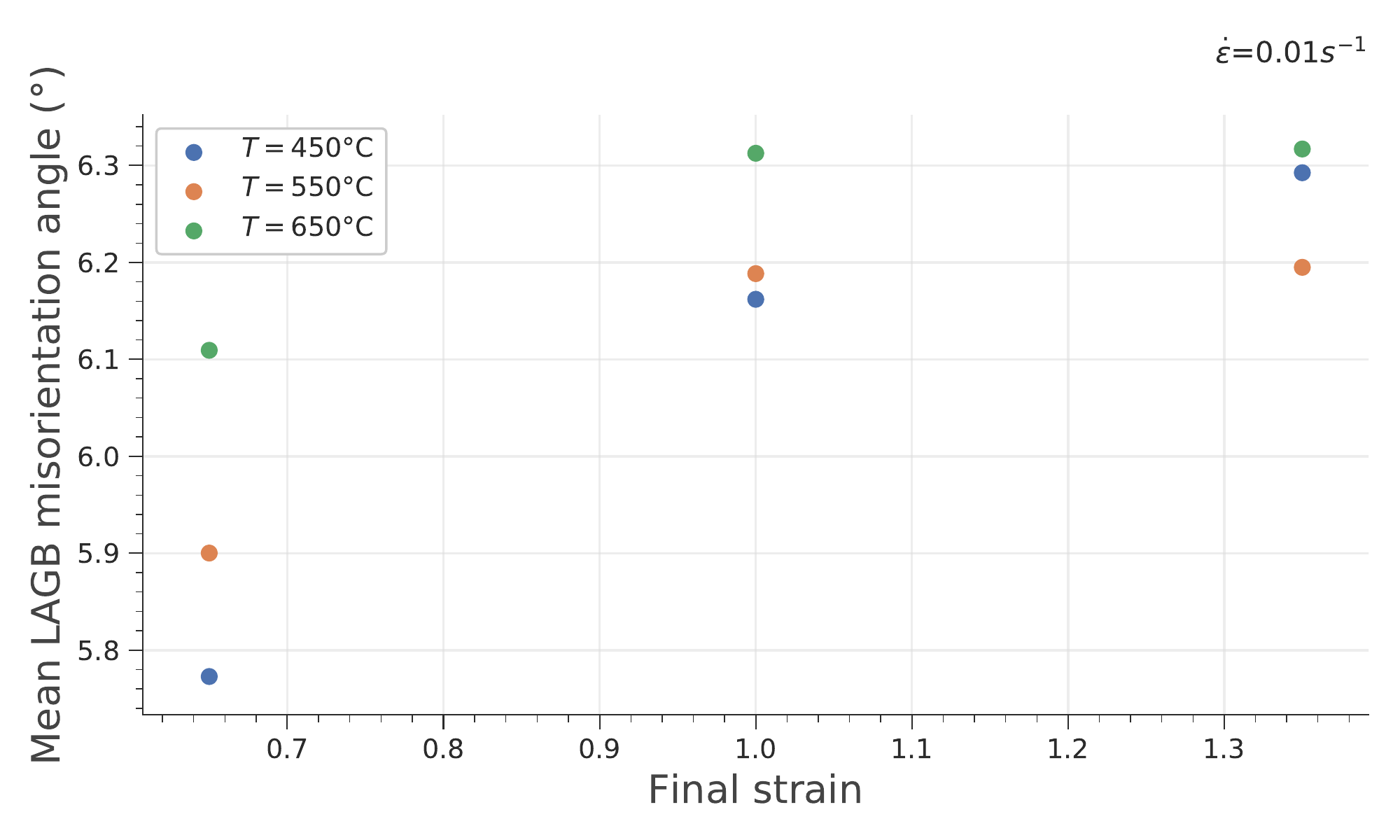}
        \caption{\label{fig:LAGBangleWithStrain} Evolution of average disorientation angle of LAGB with strain.}
    \end{subfigure}
\caption{In details investigation of DRX mechanisms.}
\label{fig:DRXmechanismsInDetails}
\end{figure}

To investigate more in details the mechanisms of DRX, some disorientation profiles are plotted as well as some average LAGB properties in figure \ref{fig:DRXmechanismsInDetails}. The disorientation profiles confirm the presence of orientation gradients, put in light some well defined LAGB with disorientation lower than $5 ^{\circ}$ and some LAGB with much higher disorientation (around $10 ^{\circ}$ and called medium-low angle GB, MLAGB). The three graphs available in figures \ref{fig:LAGBangleWithStrain}, \ref{fig:ECDWithStrain} and \ref{fig:LAGBlengthRatioWithStrain} illustrate how the substructure evolves globally in the whole microstructure as the deformation occurs and the strain increases. The high LAGB length fraction and the rather high average misorientation angle for all strain levels prove that LAGB and MLAGB populations are significant. The progressive increase of mean LAGB misorientation angle supports the hypothesis that on average, LAGB know a progressive increase of disorientation. If CDRX leads to significant evolution of the substructure, it appears that PDRX impacts to a greater extent the grain structure (see figure \ref{fig:InfluenceThermomechaConditions}).

All those elements confirm that Zy-4 undergo CDRX \cite{Zhang2021, Hadadzadeh2018} by the progressive formation of subgrains that, under subsequent deformation, will transform to recrystallized grains either by migration through a zone with orientation gradient or by stacking of dislocations \cite{Rollett2017}. They also put in light how PDRX occurs by the growth of a few recrystallized grains that sweep the deformed areas. 

\subsubsection{Influence of thermomechanical conditions}\label{subsubsec:ThermomechaConditions}

\begin{figure}
    \centering
     \begin{subfigure}{0.49\textwidth}
        \centering
	    \includegraphics[width=0.95\linewidth]{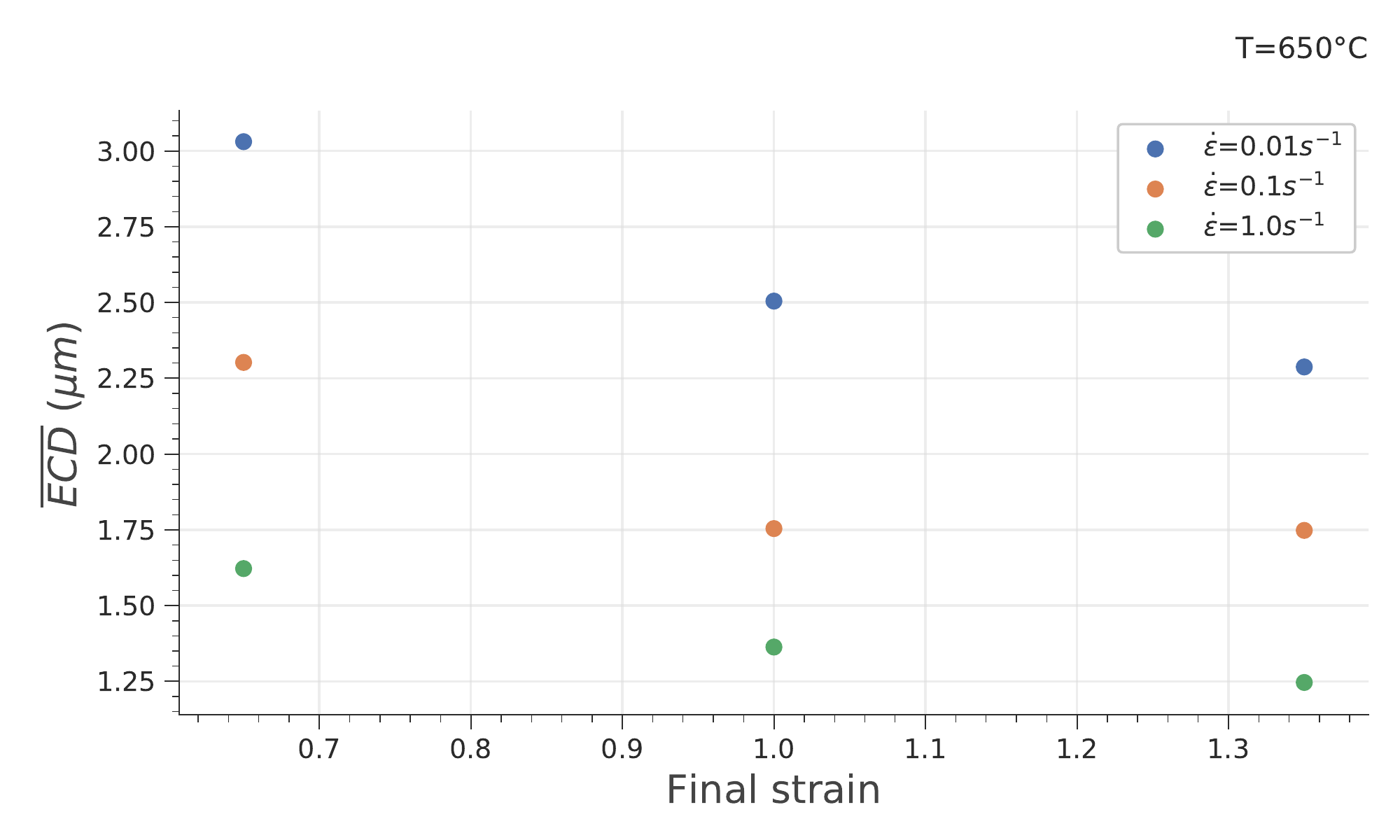}
        \caption{\label{fig:ECDWithStrain} Evolution of $\overline{ECD}$ with strain.}
    \end{subfigure}
    \begin{subfigure}{0.49\textwidth}
        \centering
        \includegraphics[width=0.95\linewidth]{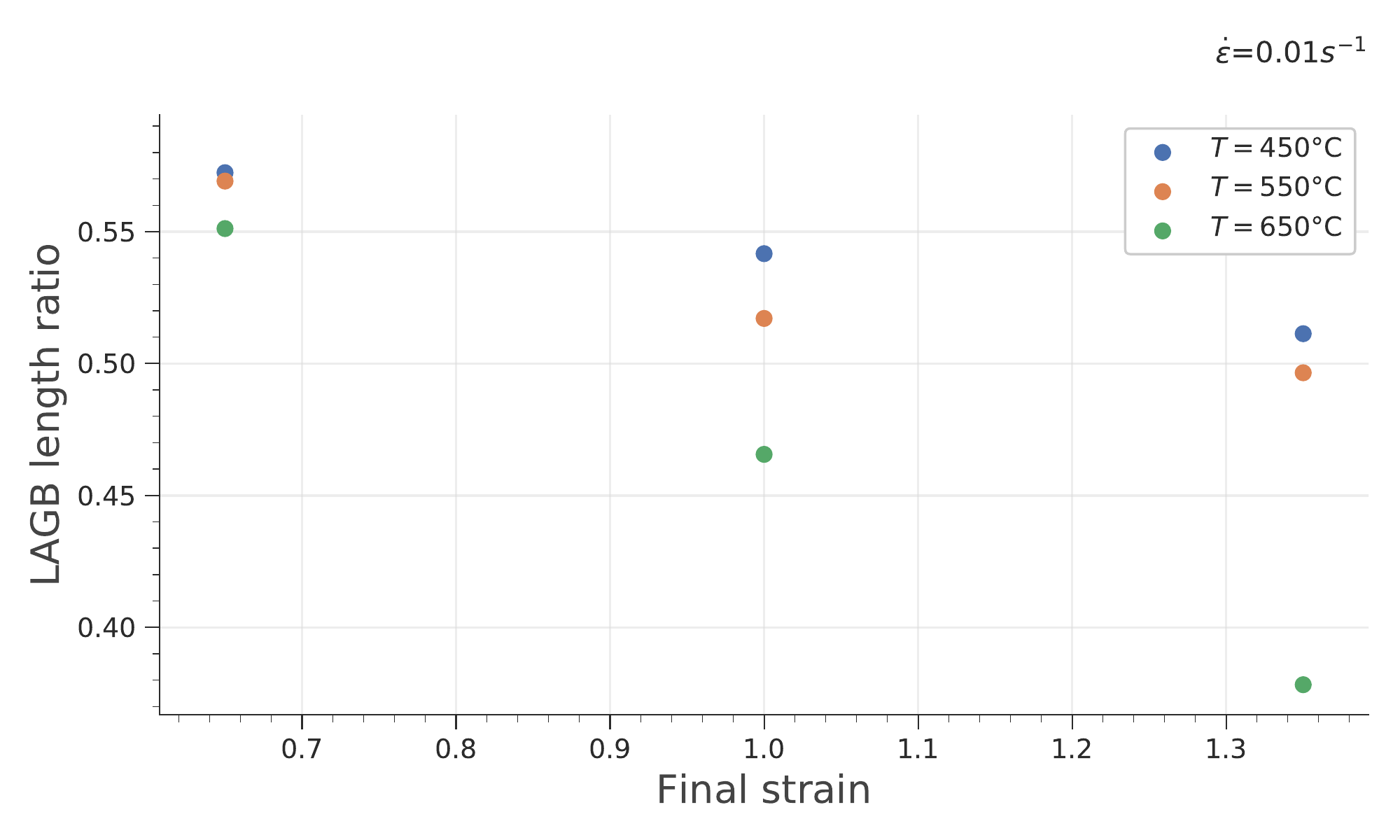}
        \caption{\label{fig:LAGBlengthRatioWithStrain} Evolution of LAGB length fraction with strain.}
    \end{subfigure}
     \begin{subfigure}{0.49\textwidth}
        \centering
	    \includegraphics[width=0.95\linewidth]{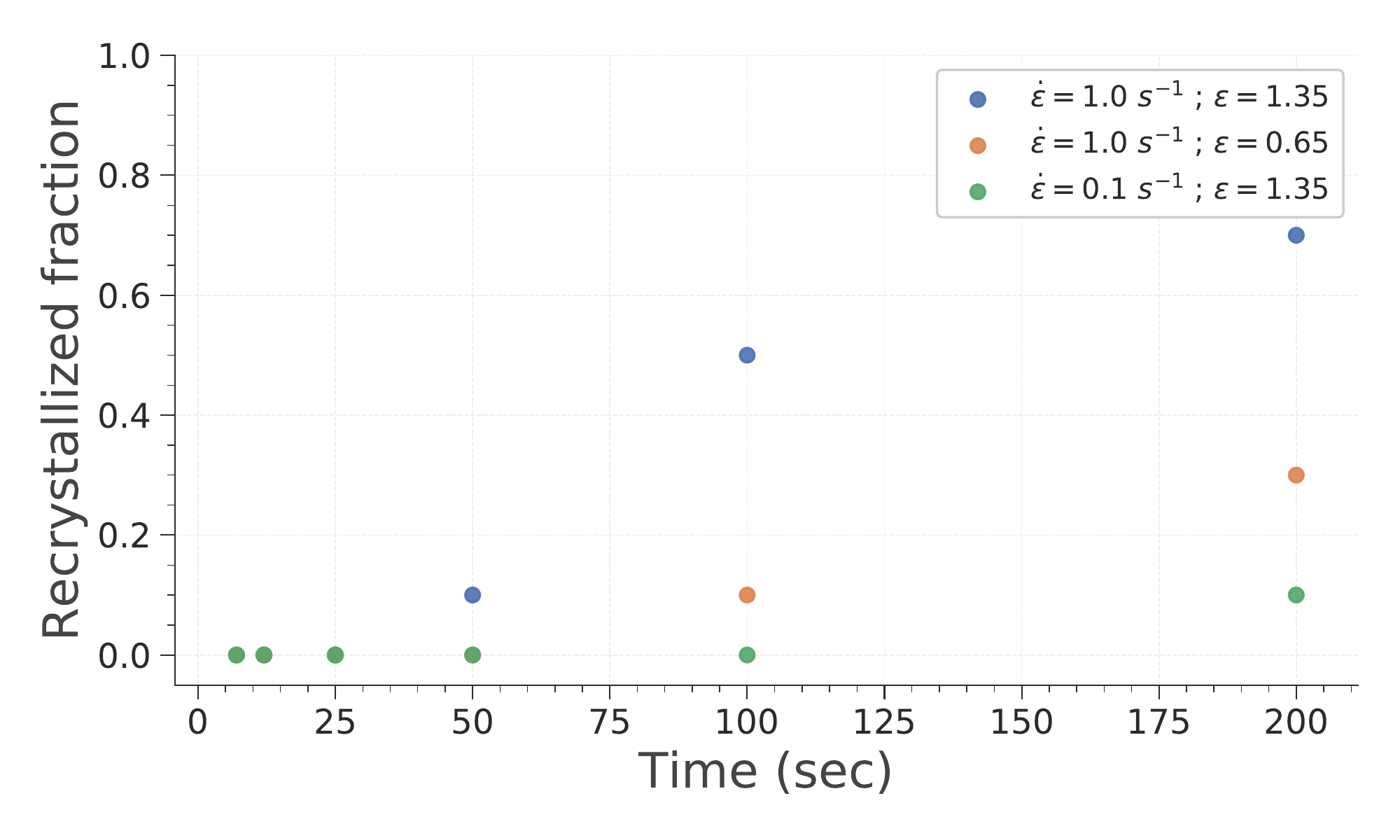}
        \caption{\label{fig:RxFractionPDRX} Evolution of recrystallized fraction with holding time.}
    \end{subfigure}
    \begin{subfigure}{0.49\textwidth}
        \centering
        \includegraphics[width=0.95\linewidth]{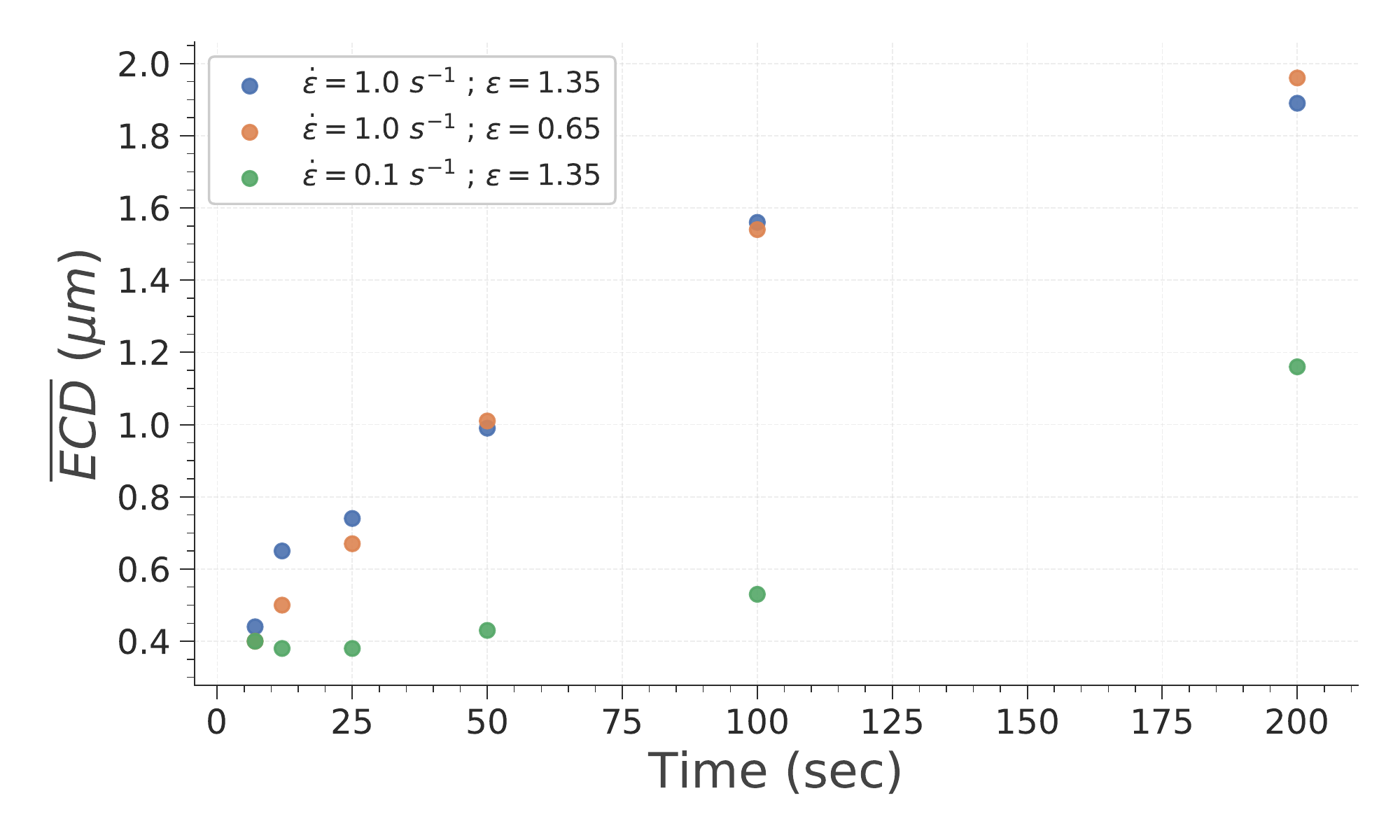}
        \caption{\label{fig:ECDPDRX} Evolution of $\overline{ECD}$ with holding time.}
    \end{subfigure}
\caption{Influence of thermomechanical conditions over some microstructure characteristic values during CDRX and PDRX.}
\label{fig:InfluenceThermomechaConditions}
\end{figure}

Orientation maps for samples quenched as fast as possible right after hot compression available on figure \ref{fig:DRXmechanisms} illustrate the wide range of observed microstructures. Qualitatively, there is not so much difference observed for the two samples deformed at $450 ^{\circ}C$ and $650 ^{\circ}C$. However, there is a significant difference for the samples deformed at strain rates of $0.01 ~ s^{-1}$ and $1.0 ~ s^{-1}$. Those qualitative observations are confirmed by the figures \ref{fig:InfluenceThermomechaConditions} and \ref{fig:LAGBangleWithStrain}. Indeed, it is clearly shown that:
\begin{itemize}
\item grain size is refined with strain and this is more significant for higher strain rates (figure \ref{fig:ECDWithStrain}). It is interesting to note that this refinement takes certainly place for low strain values and then reaches a rather steady state value.
\item Average LAGB misorientation exhibits a slight increase and illustrates the progressive disorientation of LAGB (figure \ref{fig:LAGBangleWithStrain}). This increase stays limited since some new LAGB with rather low disorientation are created upon subsequent deformation.
\item LAGB length fraction seems to decrease slightly with temperature and under subsequent deformation (figure \ref{fig:LAGBlengthRatioWithStrain}). This can be explained by the fact that LAGB starts to form quite rapidly under deformation and reaches a rather steady-state. Then, subgrain to grain transition leads to the transformation of LAGB to HAGB and to the reduction of LAGB length ratio. 
\end{itemize} 

If the samples are maintained at high temperatures for a few minutes, the influence of those thermomechanical conditions is even larger. The evolution of recrystallized fraction and mean recrystallized grain size are plotted in figure \ref{fig:InfluenceThermomechaConditions}. It illustrates how recrystallized fraction, after holding at high temperature, increases with prior deformation. The effect of strain rate upon both recrystallized fraction and recrystallized mean grain size is also put in light. Those characteristic values are not plotted for the samples quenched immediately after deformation since recrystallized fraction is found to be negligible.

\clearpage
\section{Numerical framework}\label{sec:NumericalFramework}

\subsection{Interface description}\label{subsec:InterfaceDescription}

The model used for this study relies on a level-set (LS) formulation within a finite element (FE) framework. Initial digital microstructures are generated either thanks to a Laguerre-Voronoï method \cite{Hitti2012} according to an experimental ECD distribution or by direct immersion of an EBSD image. For this second method, orientation and dislocation density are averaged inside each grain/subgrain since the model only allows for constant values inside grains. Then, interfaces are handled using LS functions initialized as signed Euclidean distance functions \cite{Bernacki2008}. Grain boundary migration is simulated by solving transport equation for each LS function:

\begin{eqnarray}
 \dfrac{\partial \phi}{\partial t} + \overrightarrow{v}\cdot \overrightarrow{\nabla\phi} = 0.
\label{eq:TransportEquation}
\end{eqnarray}

The interface velocity is supposed to be equal to the sum of a term induced by capillarity effects and of a second related to differences of stored energy between neighboring grains due to gradient of GND density \cite{Bernacki2011}:

\begin{eqnarray}
\overrightarrow{v}_c = -M \gamma \kappa \overrightarrow{n}, \label{eq:CapillarityVelocity}\\
\overrightarrow{v}_e = M \llbracket E \rrbracket \overrightarrow{n}. \label{eq:StoredEnergyVelocity}
\end{eqnarray}

$M$, $\gamma$ and $\kappa$ are respectively the mobility, energy and curvature of grain boundaries. $\llbracket E \rrbracket$ is the stored energy difference between adjacent grains and $\overrightarrow{n}$ is the outward unitary normal of grain boundaries.

Since SPP are known to reduce significantly grain boundary migration \cite{Hillert1988}, being able to take into account their effect is necessary. To do so, a population of spherical SPP respecting the experimental size distribution and surface fraction is generated. Then, mesh elements and nodes are deleted where SPP are introduced and boundary conditions are applied to replicate the interaction between grain boundaries and SPP interfaces \cite{Agnoli2014}. In case some stored energy due to deformation is considered in simulation, SPP will not be included. Indeed, pinning effect is considered to be negligible in regards to stored energy pressures. It allows to save a lot of computation resources, especially since SPP in Zy-4 are particularly fine and therefore requires to severely refine the mesh and the time step.

Different numerical improvements of this LS framework were proposed by Scholtes et al. \cite{Scholtes2016} and Maire et al. \cite{Maire2018} and are used here. 

\subsection{Subgrain formation and evolution}\label{subsec:SubgrainFormation}

To predict the formation of subgrains and their evolution under subsequent deformation, equations from Gourdet-Montheillet model \cite{Gourdet2003} have been adapted and implemented.

Dislocation density of each grain evolves according to Yoshie-Laasraoui-Jonas equation \cite{Laasraoui1991}:

\begin{eqnarray}
d\rho = \left( K_1 - K_2 \rho \right) d\varepsilon,
\label{eq:YLJ}       
\end{eqnarray}

where $K_1$ and $K_2$ are two material constants. They respectively describe the strain hardening and the recovery.

Gourdet-Montheillet model assumes that dislocations created under deformation can evolve:
\begin{itemize}
\item through their absorption by HAGB which migrate. This is naturally captured by setting to the areas swept a low dislocation density $\rho_{0}$ representing dislocation density of the material free of dislocations.
\item By organization to form subgrains. The subgrains interface created for each deformation increment ($dS^+$) by this reorganization follows next equation:
    \begin{eqnarray}
    dS^+ = \dfrac{\alpha b K_2 \rho \Dot{\varepsilon} dt}{\eta \theta_0},
    \label{eq:SurfaceNewSubgrains}       
    \end{eqnarray}
$\alpha = 1- \exp{\left( \dfrac{D}{D_0} \right) ^m}$ is a coefficient describing the fraction of dislocations recovered to form new subgrains. $D$ is the grain diameter, $D_0$ is a grain reference diameter and $m$ is a fixed coefficient. $b$ is the Burgers vector,  $\eta$ is the number of sets of dislocations and $\theta_0$ the disorientation of newly formed subgrains \cite{Gourdet2003}.  
\item By stacking into preexisting LAGB. The increment of disorientation induced by this phenomenon is modeled according to following equation \cite{Gourdet2003}:
    \begin{eqnarray}
    d\theta = \dfrac{b}{2 \eta} \left(1-\alpha\right) D K_2 \rho d\varepsilon.
    \end{eqnarray} 
\end{itemize}

Since $\alpha$ depends on grain diameter, $dS^+$ is computed individually for each grain. Thus, new subgrains naturally appear preferentially where  $dS^+$ is the highest. Subgrain diameter is initialized to respect the grain size distribution measured experimentally. Initial subgrain orientation is set by applying a small misorientation to the parent grain orientation. Misorientation axis is taken to respect uniform distribution. Misorientation angle is set to respect the disorientation measured experimentally at low strain levels. Finally, once the misorientation axis is set at subgrain formation, it is kept constant and subgrain misorientation is increased by applying a rotation around this axis.

\subsection{Material parameters}\label{subsec:MaterialParameters}

For GG simulations realized within this study, grain boundary energy and mobility are considered isotropic. This assumption is justified by the fact that the proportion of LAGB is very low in all analyzed samples and therefore the influence of anisotropy of GB properties is very limited. On the contrary, for CDRX and PDRX simulations, grain boundary energy $\gamma$ is defined by Read-Schockley equation:

\begin{eqnarray}
\gamma  (\theta) 
   	\begin{cases}
     \gamma_{max} \left( \dfrac{\theta}{\theta_{max}}\right)\left(1- \ln{\dfrac{\theta}{\theta_{max}}}\right) , ~ \theta < \theta_{max}, \\
      \gamma_{max}, ~ \theta \geq \theta_{max},
    \end{cases}
\label{eq:GammaRS}       
\end{eqnarray}

with  $\theta_{max} = 15^{\circ}$ the maximum LAGB disorientation (as for the post-process of EBSD data described in section \ref{subsec:Characterization}).

In all simulations, the mobility is considered to be isotropic and the "Heterogeneous with gradient" formulation is employed \cite{Murgas2021}. A discussion about misorientation dependence of mobility is discussed in \cite{Murgas2021, Grand2022}.

$K_2$ is supposed to be constant for all grains. To reproduce intergranular deformation heterogeneity, $K_1$ varies from grain to grain according to a distribution. Taylor equation is used to deduce flow stress from average dislocation density. Thus, $\overline{K_1}$ and $K_2$ are identified based on the stress-strain curves and on the GND density values measured using EBSD. To determine $K_1$ distribution, the assumption is made that for strain values around $1.35$, all grains have reached their saturation GND density ($\rho_{sat}$). Using equation \ref{eq:YLJ}, one can deduce that:

\begin{eqnarray}
 \rho_{sat} = \lim_{\varepsilon\to\infty} rho = \dfrac{K_1}{K_2}.
\label{eq:RhoSaturation}       
\end{eqnarray}

Accordingly, one can get from GND density estimated by EBSD the value of $K_1$ for each individual grain and deduce its statistical distribution.

Finally, GND density is converted to energy using next equation: 

\begin{eqnarray}
E = \tau \rho,
\end{eqnarray}

where $\tau$ stands for dislocation line energy and is estimated through relation: $\tau = \dfrac{\mu b^2}{2}$ with $\mu$ the shear modulus \cite{Dieter1976}.

An extensive description of the numerical framework can be found in literature \cite{Bernacki2011, Murgas2021, Scholtes2015, Maire2017}. Developments dedicated to CDRX modeling will be detailed in an upcoming article \cite{Grand2022}.

\section{Modeling of grain growth and recrystallization}\label{sec:Model}

\subsection{Modeling of grain growth}\label{subsec:ModelGG}

As it has been done regarding experimental characterization of GG, this section first focuses on data obtained with the material having no SPP. Then, identified parameters are used to simulate GG in the presence of SPP.

\subsubsection{Identification of material parameters without SPP}\label{subsubsec:WithoutSPP}

In pure GG context, there is no stored energy, so the term $v_e$ is null and the interface velocity is simply equal to the capillarity term $v_c$ (see equations \ref{eq:CapillarityVelocity} and \ref{eq:StoredEnergyVelocity}). Therefore, for being able to run GG simulations, one should have values for the product $M \times \gamma$, also denoted as reduced mobility. Using experimental data presented in section \ref{subsec:CharacGG}, it is possible to get initial values for reduced mobility by fitting a Burke and Turnbull law \cite{Burke1952}. This allows to run a first set of simulation. Then, comparing simulation results to experimental data, it is possible to adjust the reduced mobility value by a given coefficient in order to minimize $L^2$ error. Finally, a last set of simulations is launched using this adjusted reduced mobility values. This succession of operations is of course repeated for all temperatures, in our cases $600$, $650$ and $700 ^{\circ}C$.
The last step of the identification procedure consists in interpolating, and if required, extrapolating, the reduced mobility parameters identified at different temperatures.  To do so, $\gamma$ is considered to be constant within the temperature range and is taken from literature \cite{Dunlop2007_Rx}. Then, an Arrhenius law is fitted such as described by equation \ref{eq:ArrheniusReducedMobility}. This procedure is also applied and described in details in ref. \cite{Alvarado2021_Exp}.

\begin{eqnarray}
M = M_0 \times \exp{\dfrac{-Q}{RT}}.
\label{eq:ArrheniusReducedMobility}
\end{eqnarray}

In equation \ref{eq:ArrheniusReducedMobility}, $M_0$ is considered as a constant, $Q$ is the activation energy for GG and $R$ is the gas constant.

Figure \ref{fig:IdReducedMobility} presents the evolution of $\overline{ECD}$ for the three temperatures. One can note that the general agreement between experimental and numerical results is satisfying. Few differences are still observable, particularly for very short and very long heat treatments. Indeed, the kinetics seems to be slightly underestimated for short durations and slightly overestimated for long times. However, the simulation results still show about less than $1 ~ \mu m$ difference compared to experimental results which remains precise enough.

\subsubsection{Application to commercial material}\label{subsubsec:WithSPP}

\begin{figure}
    \centering
    \begin{subfigure}{0.45\textwidth}
        \centering
        \includegraphics[width=0.95\linewidth]{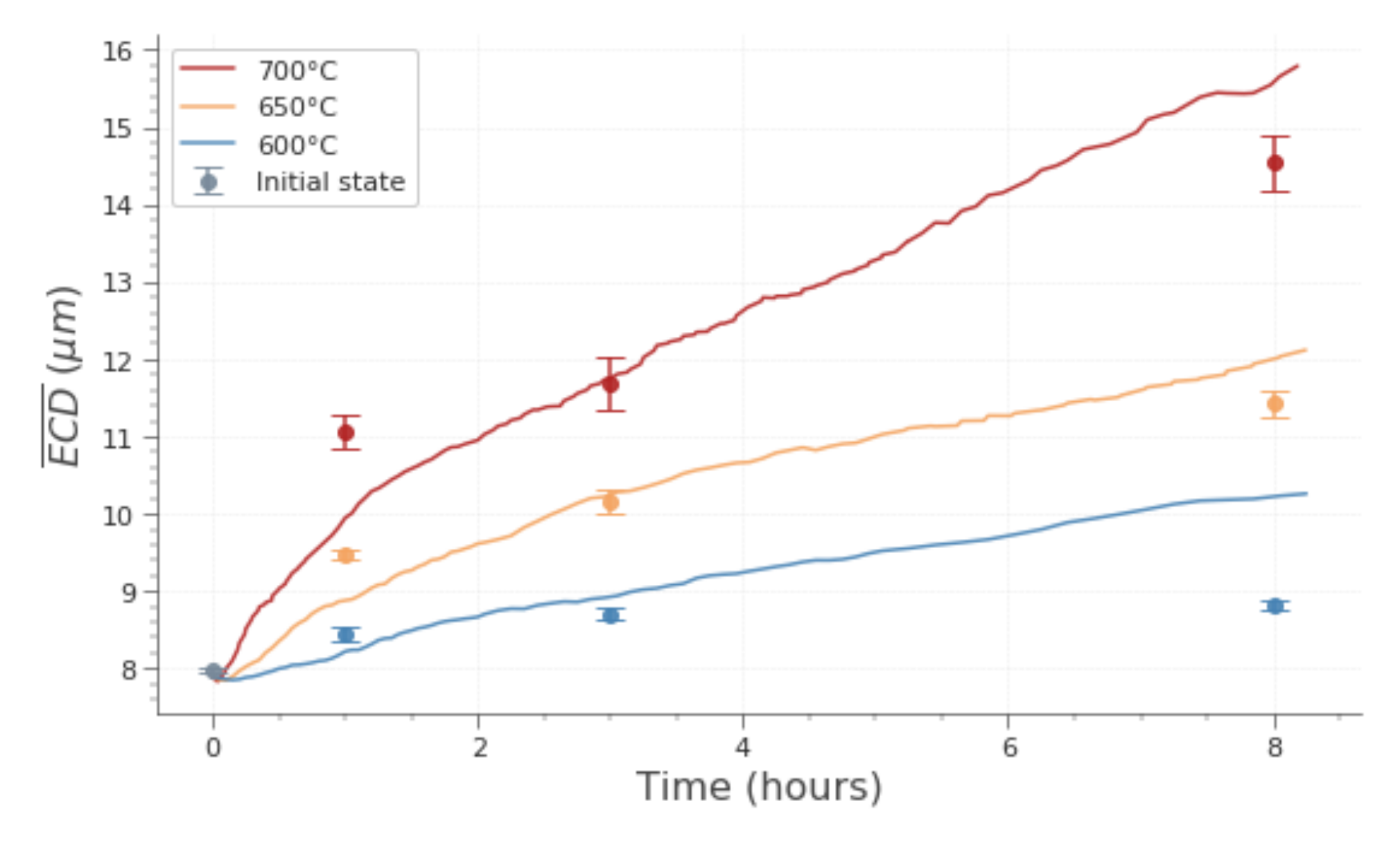}
        \caption{\label{fig:IdReducedMobility} Evolution of $\overline{ECD}$ with time.}
    \end{subfigure}
    \begin{subfigure}{0.45\textwidth}
        \centering
        \includegraphics[width=0.95\linewidth]{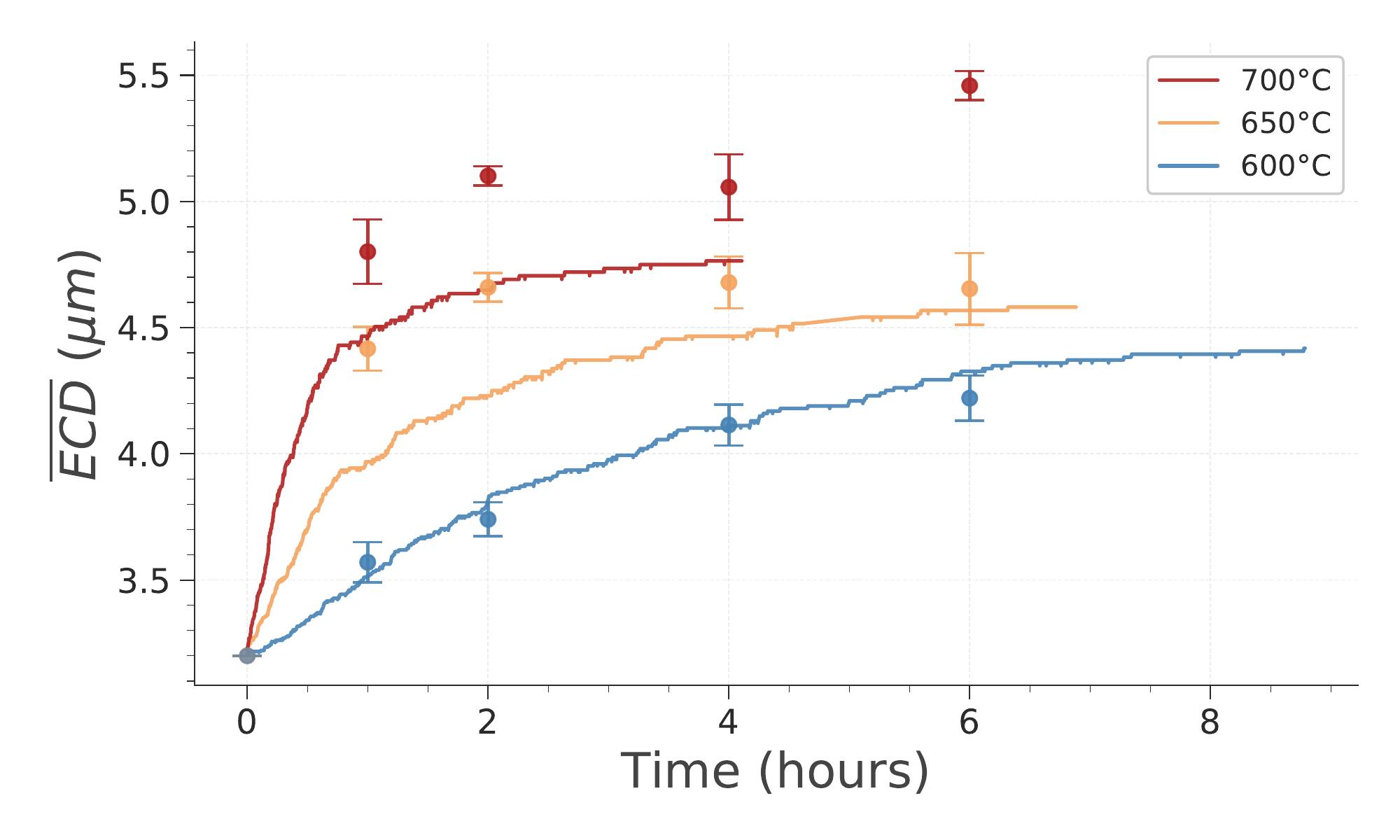}
        \caption{\label{fig:ECDFunctionOfTimeGGSPP} Evolution of $\overline{ECD}$ with time.}
    \end{subfigure}
    \begin{subfigure}{0.32\textwidth}
        \centering
        \includegraphics[width=0.98\linewidth]{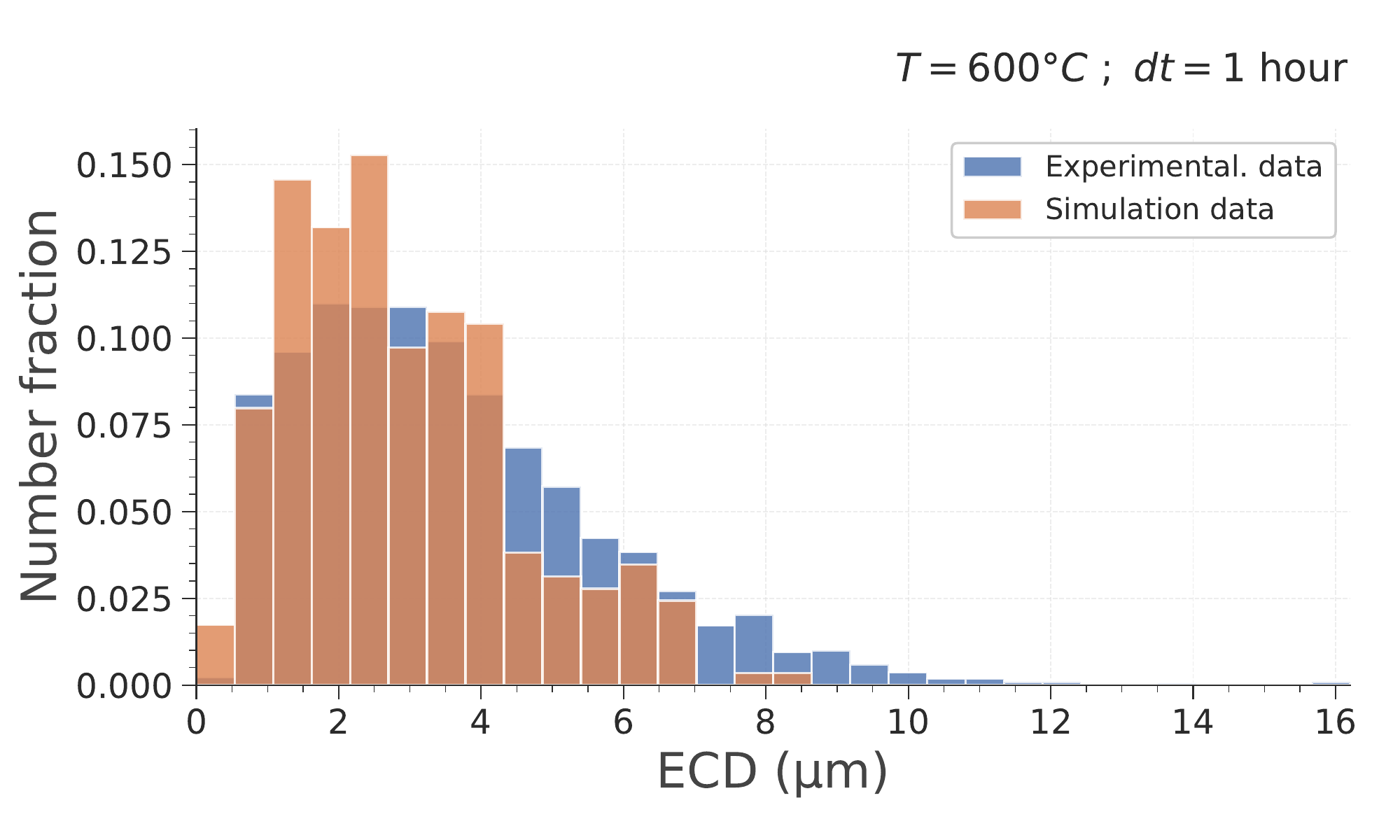}
        \caption{\label{fig:Hist1ECDGGSPP} $ECD$ number histogram at $t= 1 ~ hour$.}
    \end{subfigure}
    \begin{subfigure}{0.32\textwidth}
        \centering
        \includegraphics[width=0.98\linewidth]{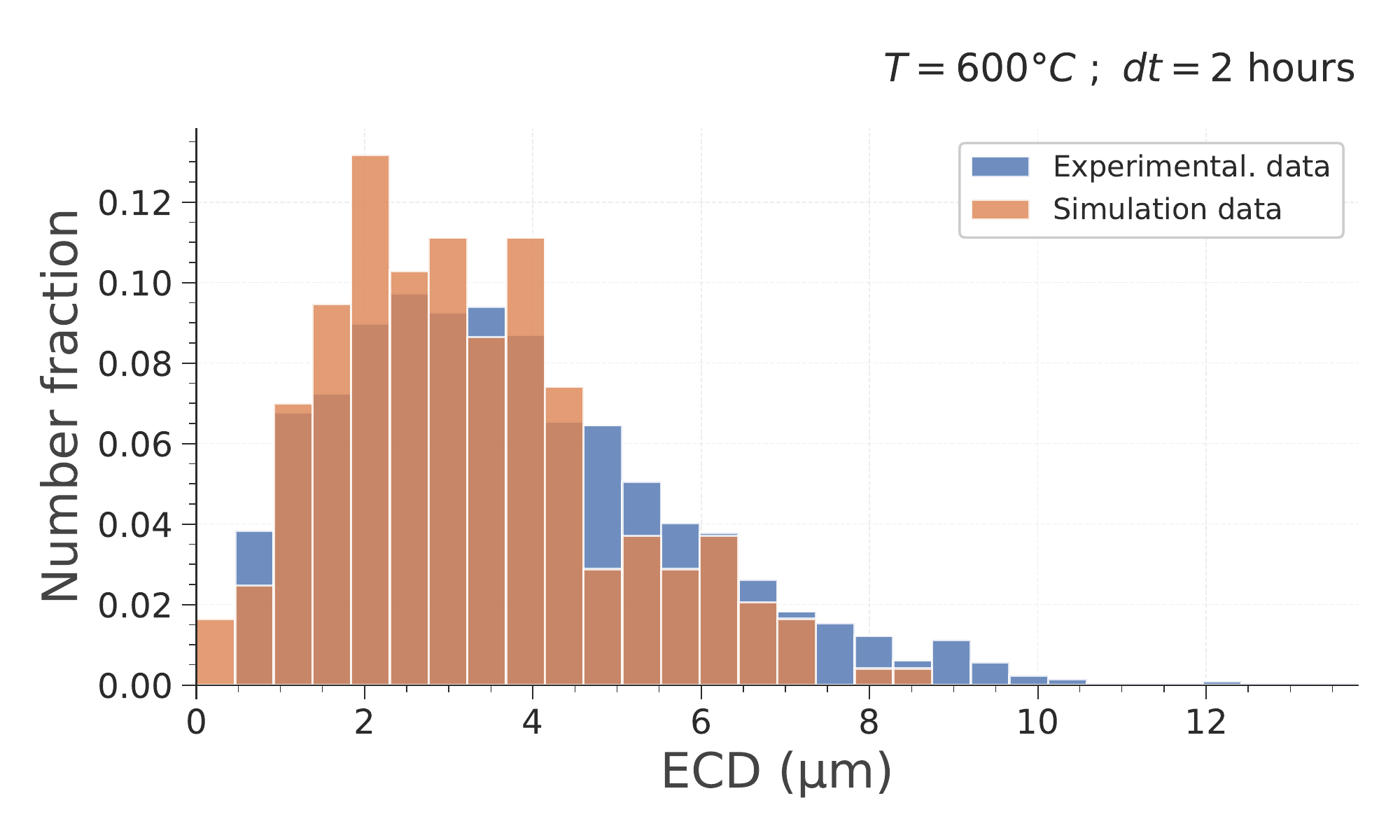}
        \caption{\label{fig:Hist2ECDGGSPP} $ECD$ number histogram at $t= 2 ~ hours$.}
    \end{subfigure}
        \begin{subfigure}{0.32\textwidth}
        \centering
        \includegraphics[width=0.98\linewidth]{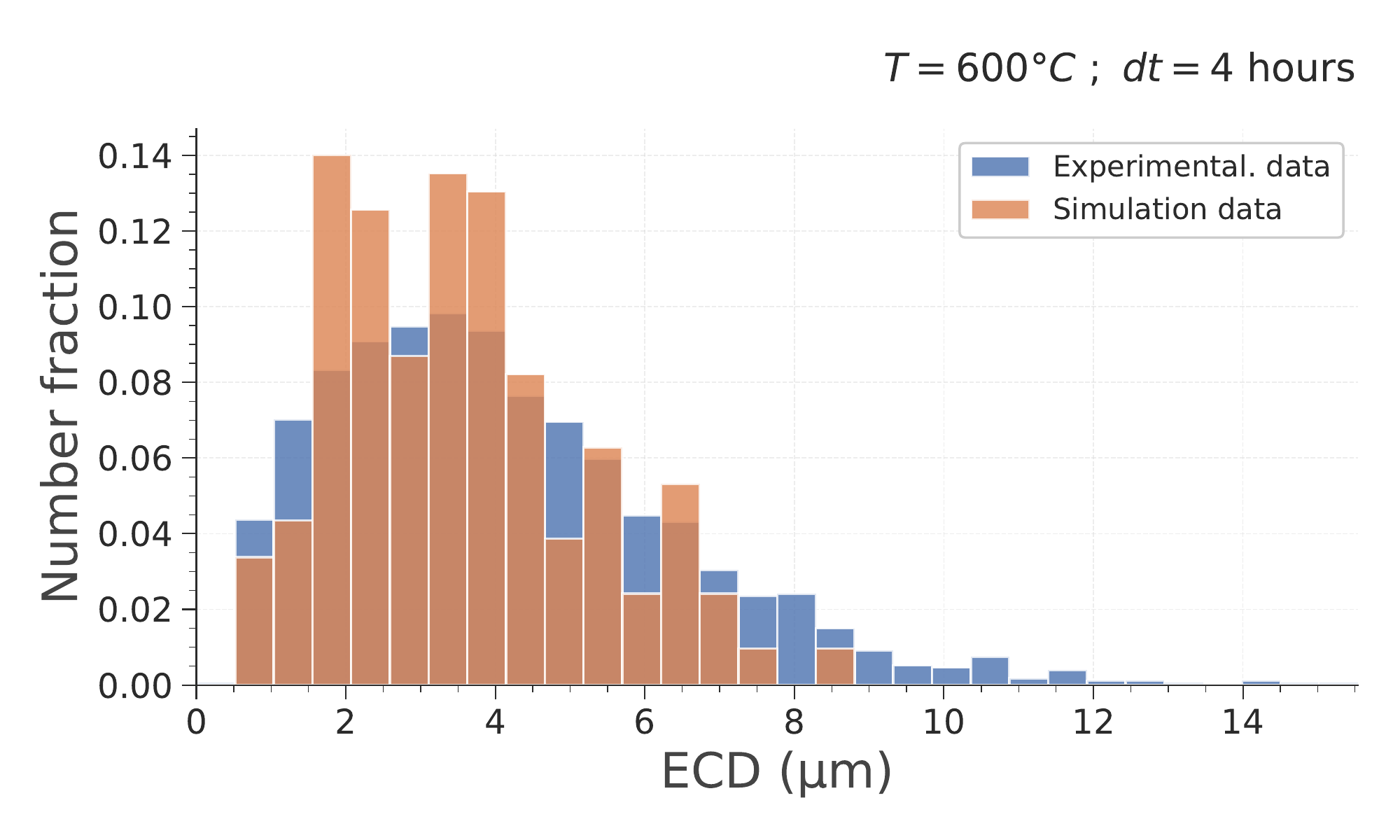}
        \caption{\label{fig:Hist3ECDGGSPP} $ECD$ number histogram at $t= 4 ~ hours$.}
    \end{subfigure}
\caption{Evolution of grain size population during GG simulation.}
\label{fig:GGSimulation}
\end{figure}

After having identified reduced mobility using data representative of the material presenting no SPP, it is particularly interesting to assess whether these parameters are appropriate for modeling GG with a significant SPP population. Simulation results detailed within this section follow the same principles than the ones described in section \ref{subsubsec:WithoutSPP}. SPP populations are included within this simulation according to the strategy described in section \ref{subsec:InterfaceDescription}. SPP are considered to be static within the heat treatment. Nevertheless, to ensure maximal representativeness, SPP populations are defined independently for each heat treatment temperature according to the experimental measurements done by TEM and describing the state of SPP at the end of the heat treatment (see section \ref{subsec:CharacGG}). Figure \ref{fig:GGSimulation} presents the evolution of the $\overline{ECD}$ with time as well as $ECD$ histograms at different instants.

The description of mean grain size and of the general grain size distribution is excellent for the lowest temperature. Moreover, the simulations are able to reproduce the fact that $\overline{ECD}$ plateau at three distinct values. According to classic Smith-Zener theory of GB pinning by SPP, $\overline{ECD}_{sat} \propto R_{\textsc{spp}}/f_{\textsc{spp}}$. Therefore, this increase of $\overline{ECD}_{sat}$ with temperature is consistent with a phenomenon of Ostwald ripening happening during heat treatment. Nevertheless, numerical results underestimate GG at $700^{\circ} C$. This could be explained by the fact that TEM measurements of SPP population are made at room temperature and so the SPP populations considered for these simulations are potentially not fully representative due to some dissolution of SPP. Finally, if one had access to the kinetics of SPP population evolution, it would be possible to use the latest developments made in the used LS framework to consider a population of SPP evolving with time \cite{Alvarado2021}.

\subsection{Modeling of CDRX and PDRX}\label{subsec:ModelDRX}

\begin{figure}[h!]
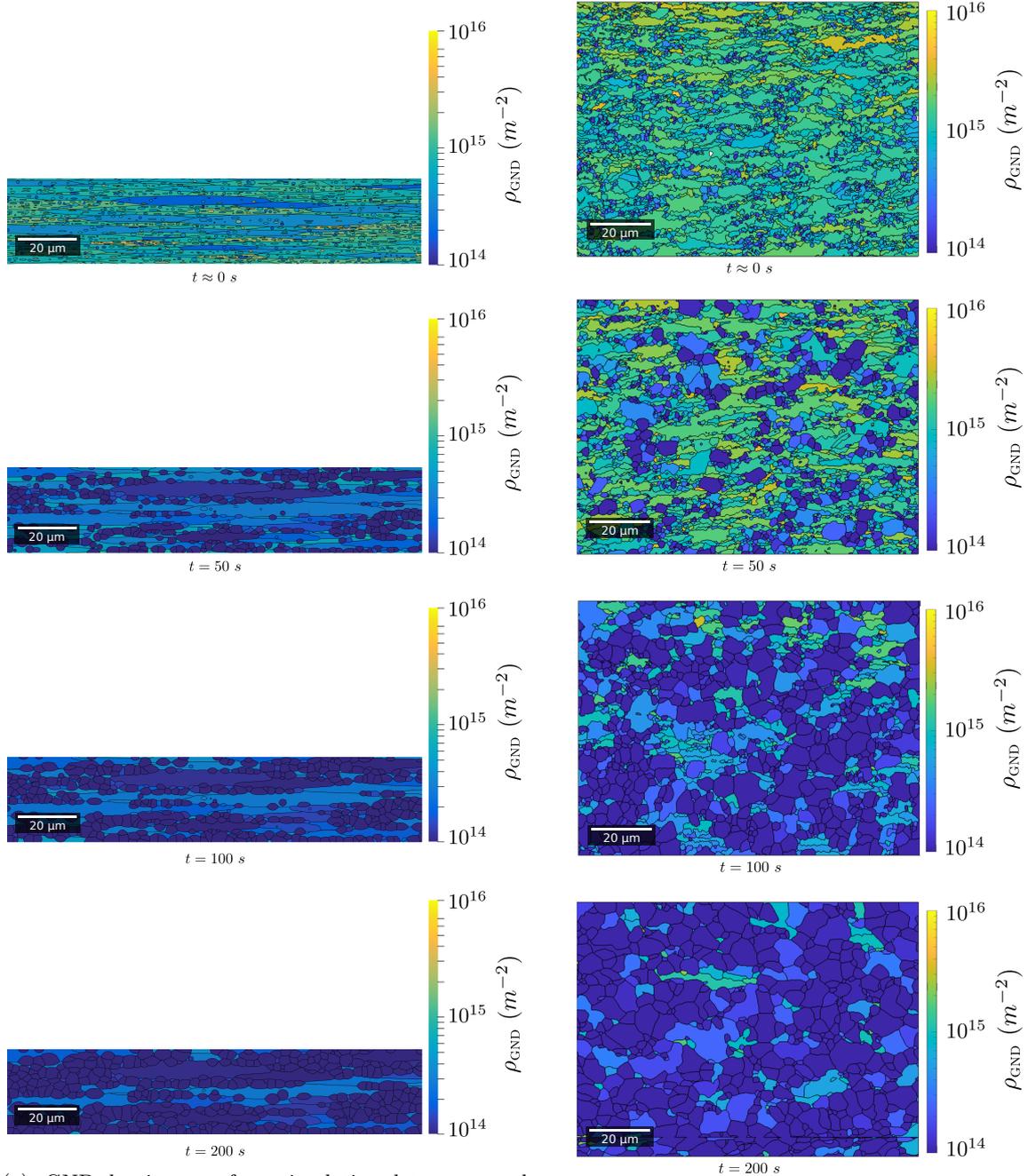

    \centering
    \begin{subfigure}[h!]{0.49\textwidth}
        \centering
        \includegraphics[width=0.98\linewidth]{figures/CDRX_GNDdensity_SimuMicrostructures_B_C.pdf}
        \caption{\label{fig:GNDDensityNoLAGBSimuMicroCDRX} GND density map from simulation data, averaged by subgrain.}
    \end{subfigure}
    \begin{subfigure}[h!]{0.49\textwidth}
        \centering
        \includegraphics[width=0.85\linewidth]{figures/CDRX_GNDdensity_ExpMicrostructures.pdf}
        \caption{\label{fig:GNDDensityNoLAGBExpMicroCDRX} GND density map computed from experimental data, averaged by grain.}
    \end{subfigure}
\caption{Comparison of experimental and digital GND density maps during CDRX and PDRX.}
\label{fig:NumVsExpMicroCDRX}
\end{figure}

The implementations dedicated to CDRX simulation described in section \ref{subsec:SubgrainFormation} are applied to propose a representative simulation of both CDRX and PDRX, starting from a material with an equiaxed microstructure. Initial microstructure is generated in order to respect experimental grain size and orientation distribution representative of the initial state. Then, this representative elementary volume (REV) is deformed until $\varepsilon = 1.35$ at $650 ^{\circ}$ and is hold for $300 s$. Some numerical and experimental GND density maps are displayed in figure \ref{fig:NumVsExpMicroCDRX}. The general qualitative agreement between experimental and numerical microstructures is good. It is noteworthy to mention that the model is able to capture microstructure heterogeneities with preferential zones for the formation and growth of recrystallized grains. Indeed, as shown in figure \ref{fig:GNDDensityNoLAGBSimuMicroCDRX}, the model is able to reproduce to some extent intergranular deformation heterogeneities. Therefore, the grains having a higher dislocation density are the first ones to be swept by recrystallized grains. 

\begin{figure}[h!]
    \centering
    \begin{subfigure}{0.45\textwidth}
        \centering
        \includegraphics[width=0.95\linewidth]{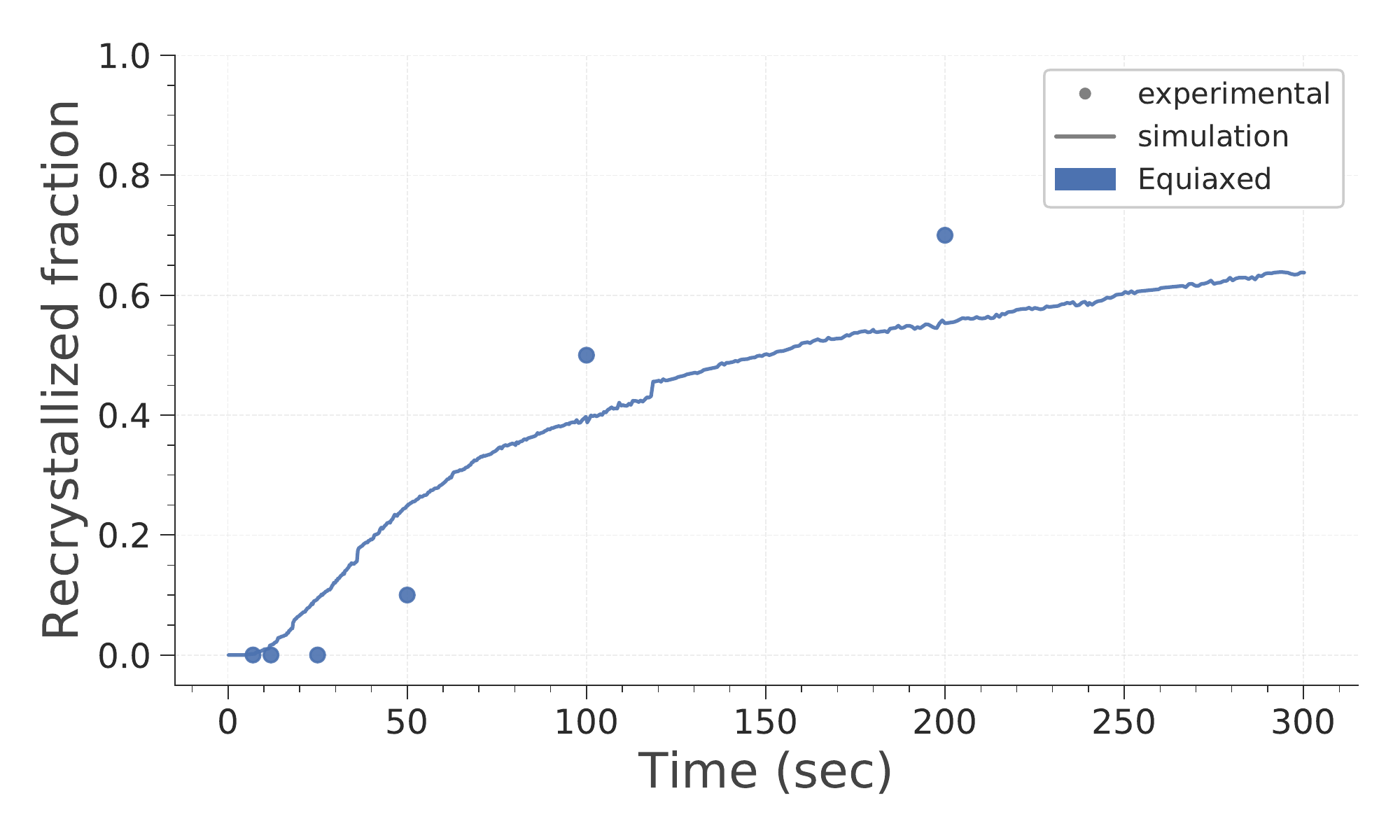}
        \caption{\label{fig:RxFractionFunctionOfTimeCDRX} Evolution of recrystallized fraction with time.}
    \end{subfigure}
    \begin{subfigure}{0.45\textwidth}
        \centering
        \includegraphics[width=0.95\linewidth]{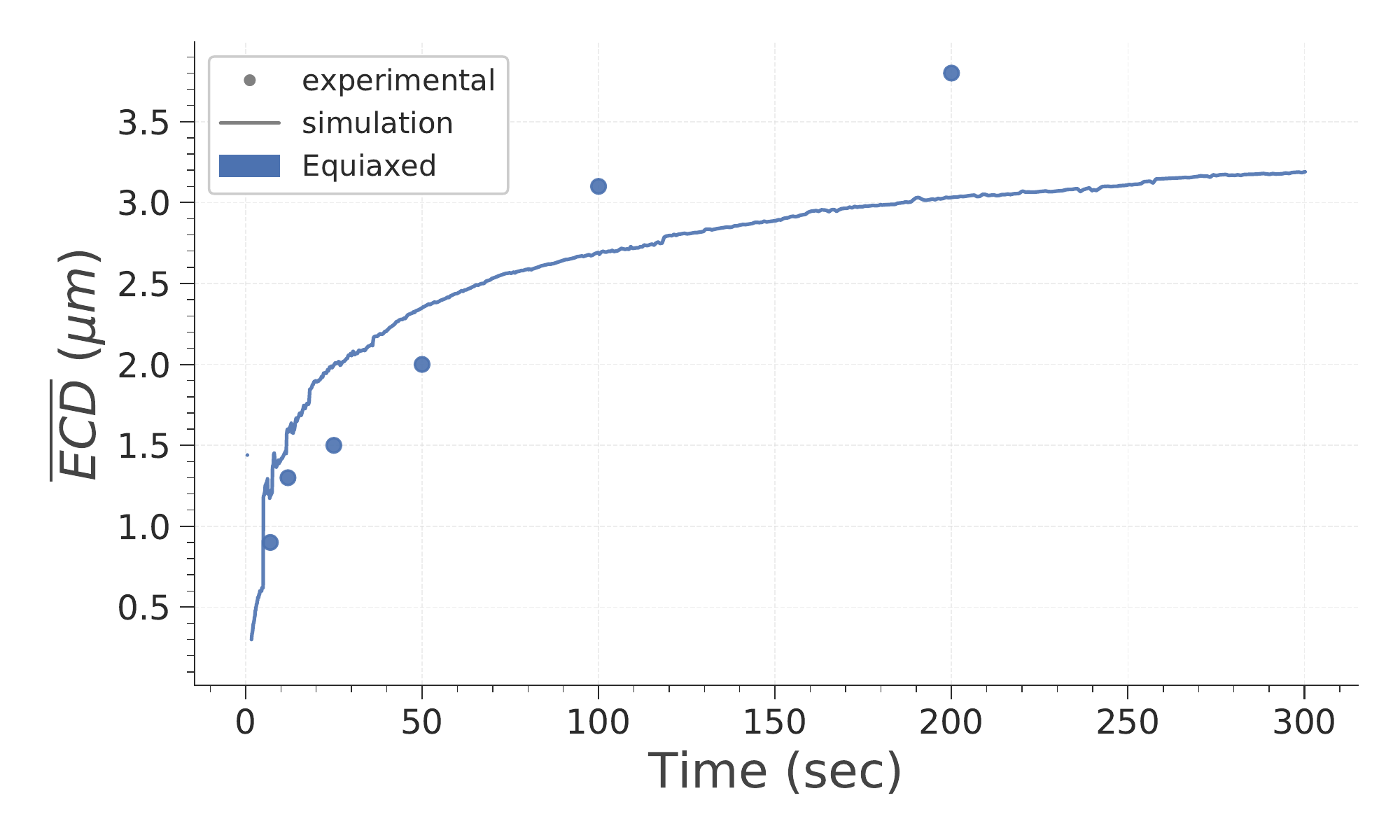}
        \caption{\label{fig:ECDFunctionOfTimeCDRX} Evolution of RX grain $\overline{ECD}$ with time.}
    \end{subfigure}
    \begin{subfigure}{0.45\textwidth}
        \centering
        \includegraphics[width=0.95\linewidth]{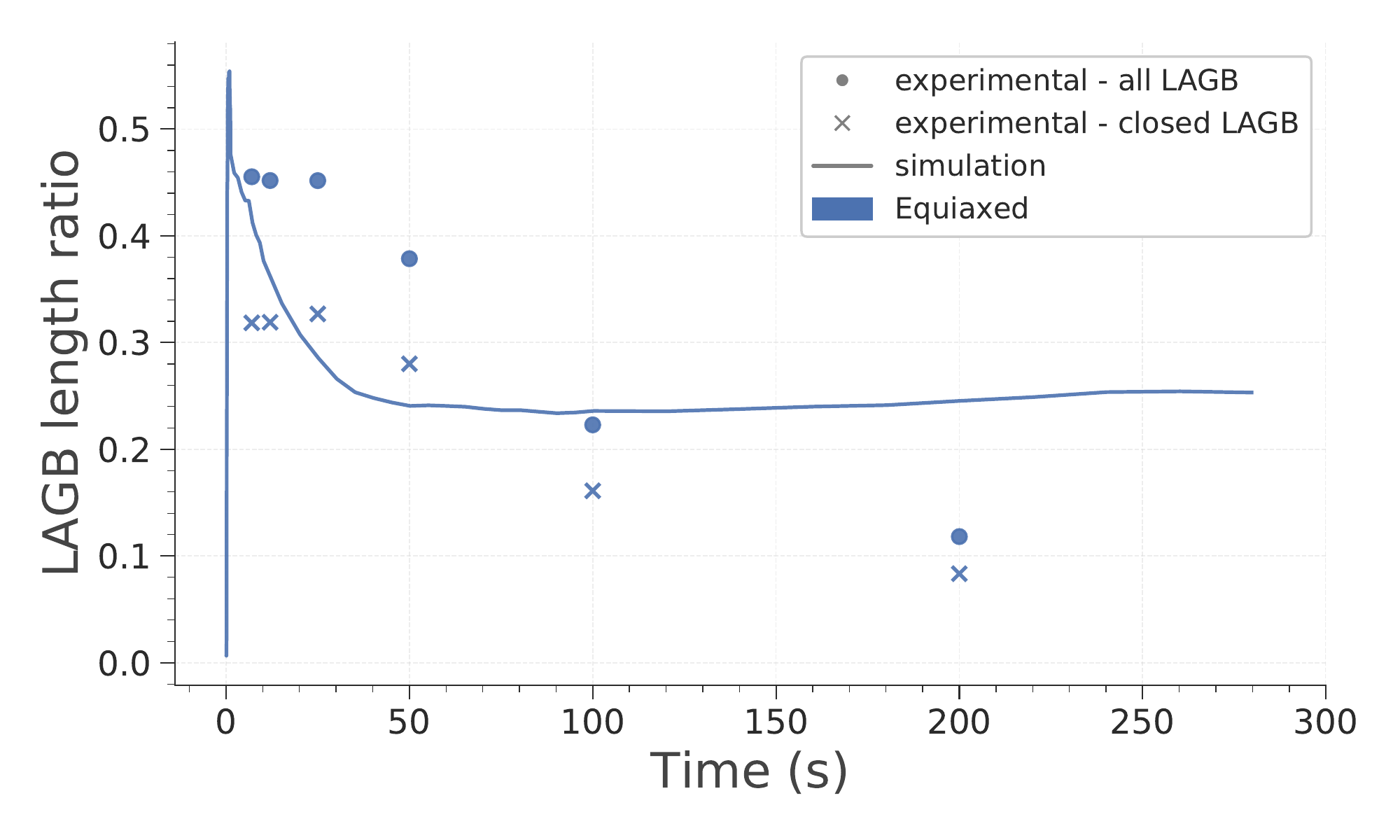}
        \caption{\label{fig:LAGBFractionFunctionOfTimeCDRX} Evolution of LAGB length fraction with time.}
    \end{subfigure}
        \begin{subfigure}{0.45\textwidth}
        \centering
        \includegraphics[width=0.95\linewidth]{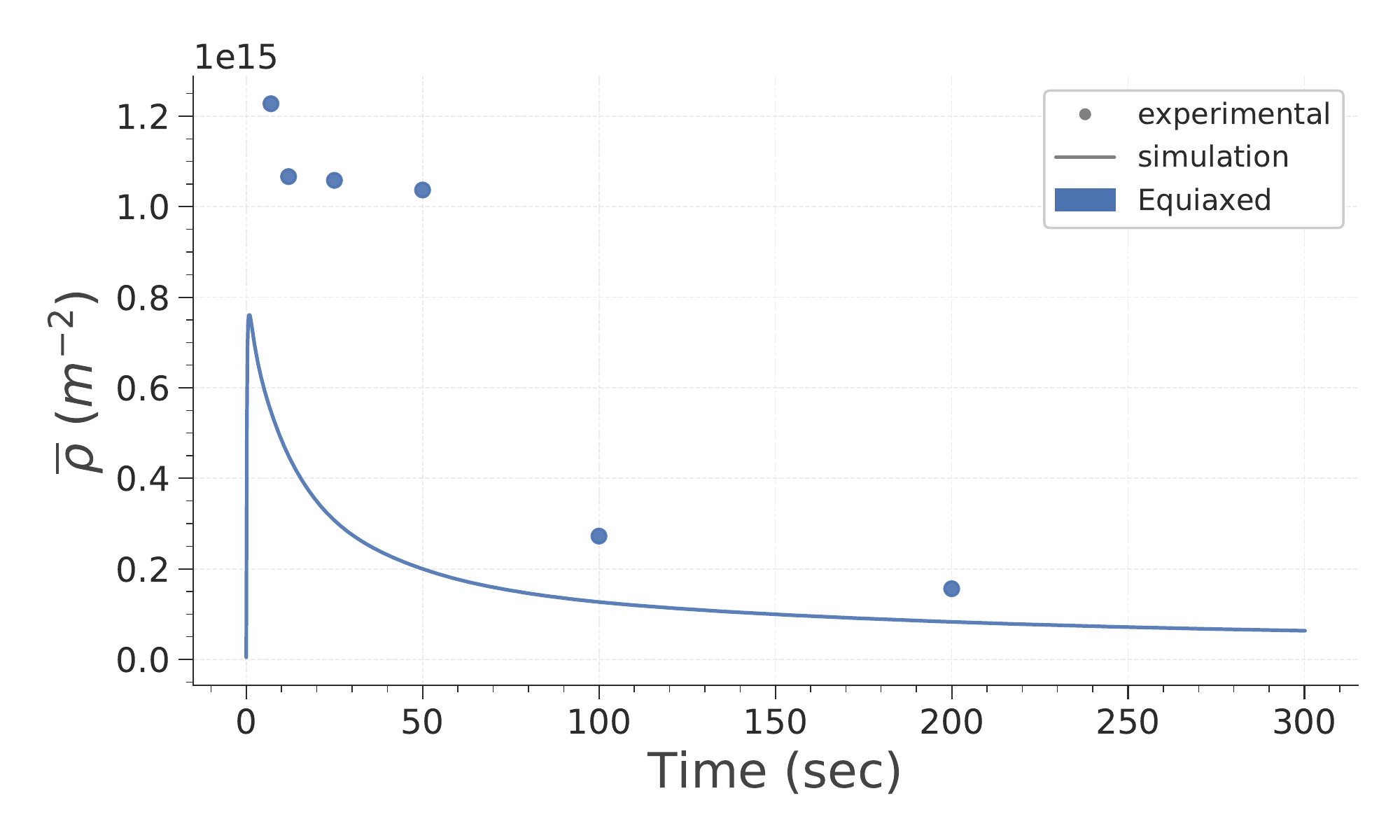}
        \caption{\label{fig:RhoFunctionOfTimeCDRX} Evolution of $\overline{\rho}$ with time.}
    \end{subfigure}
\caption{Evolution of the main microstructural features during hot deformation followed by heat treatment simulation.}
\label{fig:NumVsExpGraphsCDRX}
\end{figure}

Finally, a quantitative comparison of some characteristic values is presented in figure \ref{fig:NumVsExpGraphsCDRX}. Recrystallized fraction and recrystallized $\overline{ECD}$ are well predicted by the model. The model is able to evaluate correctly the length of LAGB created during hot deformation and is in the right order of magnitude during subsequent heat treatment. Nevertheless, it is not able to describe properly the first instants of PDRX during which LAGB length ratio stays constant.

\subsection{Modeling of PDRX using immersed EBSD data}\label{subsec:ModelPDRX}

The most significant microstructure evolutions caused by recrystallization take place during heat treatment after hot deformation (see section \ref{subsubsec:RxMechanism}). Therefore, to study in details the influence of the initial microstructure, it appears interesting to use the EBSD data to define precisely the initial microstructure. This is particularly relevant since with the actual framework, deformation phenomena are simplified due to the fact that modeling deformation using state of the art models such as crystal plasticity is too costly \cite{RuizSarrazola2020}.

\begin{figure}[h!]
    \centering
    \begin{subfigure}{0.9\textwidth}
        \centering
        \includegraphics[width=0.95\linewidth]{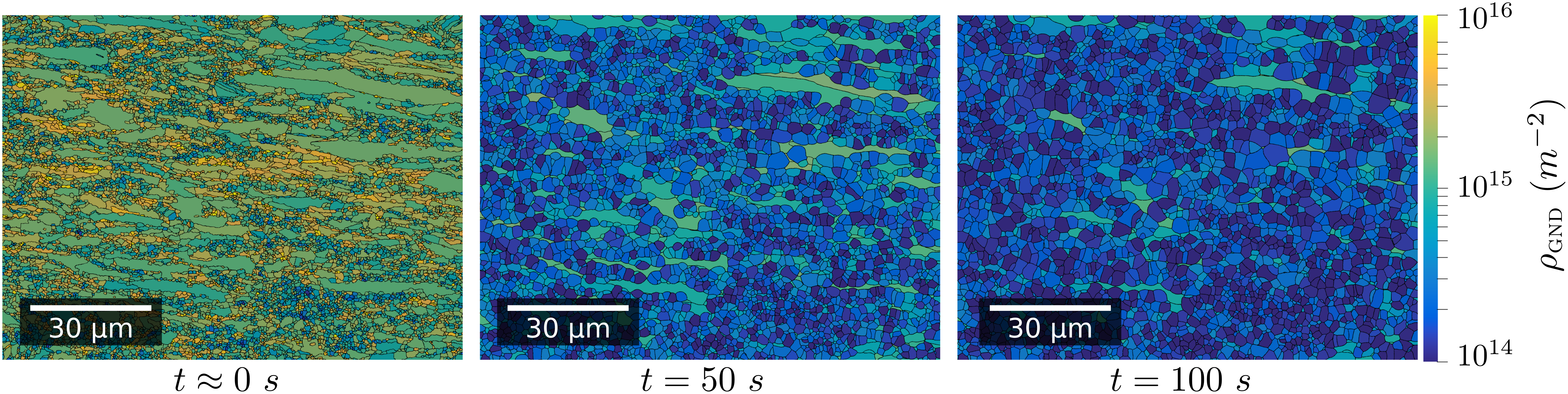}
        \caption{\label{fig:GNDDensityNoLAGBSimuMicroPDRX} GND density map from simulation data, averaged by subgrain.}
    \end{subfigure}
    \begin{subfigure}{0.9\textwidth}
        \centering
        \includegraphics[width=0.95\linewidth]{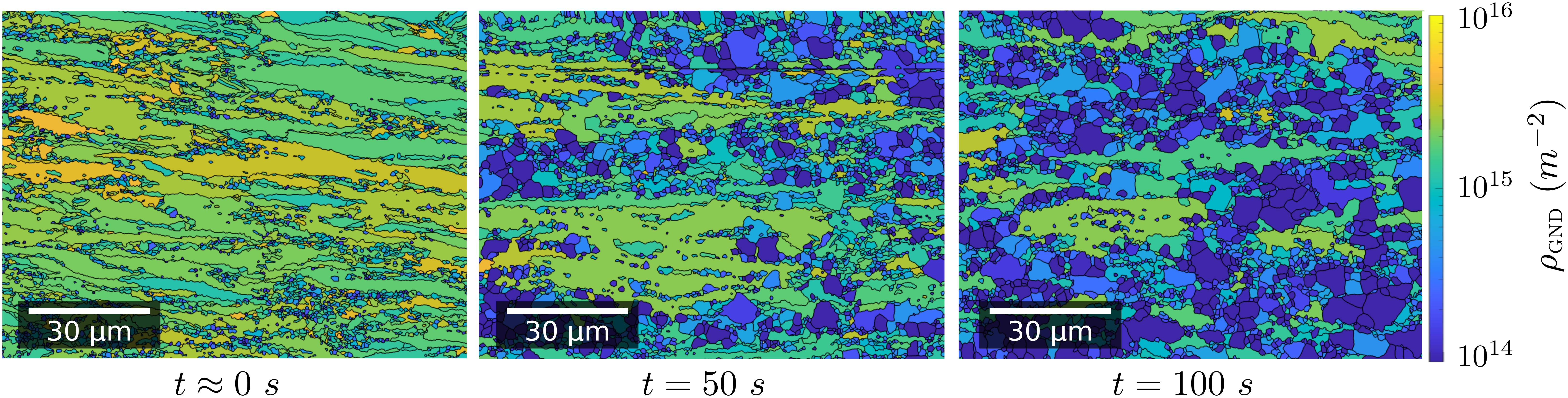}
        \caption{\label{fig:GNDDensityExpMicroPDRX} GND density map computed from experimental data, average by grain.}
    \end{subfigure}
    \begin{subfigure}{0.9\textwidth}
        \centering
        \includegraphics[width=0.95\linewidth]{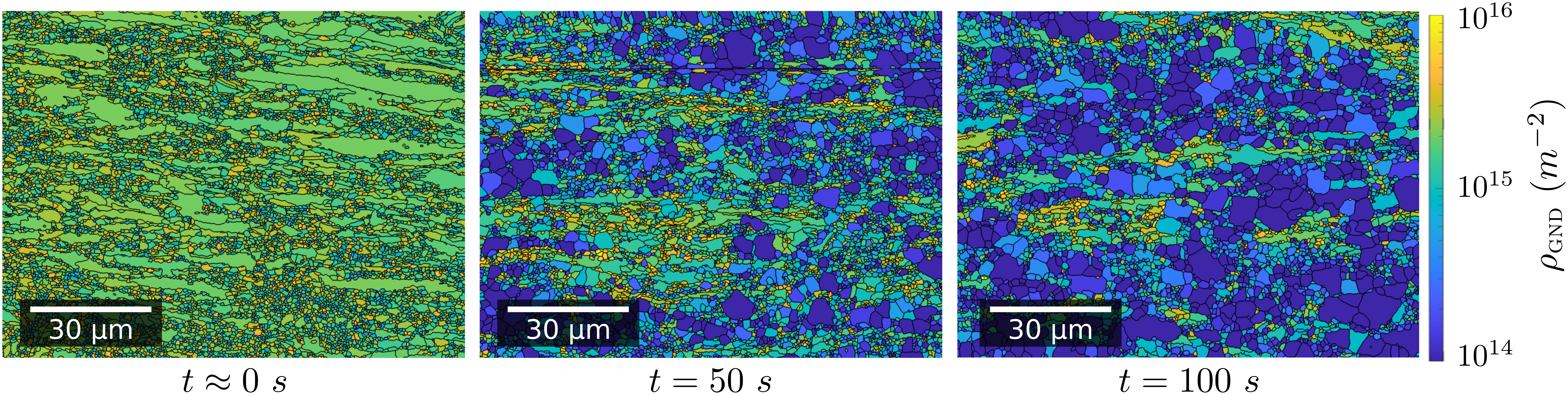}
        \caption{\label{fig:GNDDensityNoLAGBExpMicroPDRX} GND density map computed from experimental data, averaged by grain/subgrain.}
    \end{subfigure}
\caption{Comparison of experimental and digital GND density maps during PDRX.}
\label{fig:NumVsExpMicroPDRX}
\end{figure}

EBSD data from two samples are post-processed as described in section \ref{subsec:Characterization}. Those two samples have been deformed under the same thermomechanical conditions: $T = 650 ^{\circ}C ~ ; ~ \dot{\varepsilon} = 1.0 ~ s^{-1} ~ ; ~ \varepsilon_f = 1.35$. Figure \ref{fig:NumVsExpMicroPDRX}  illustrates how both digital and experimental microstructures evolve during heat treatment. One can see that the exact same EBSD data, post-processed either by averaging GND density into grains (figure \ref{fig:GNDDensityExpMicroPDRX}) or subgrains (figure \ref{fig:GNDDensityNoLAGBExpMicroPDRX}) can influence drastically the general look of the microstructure. Then, comparing experimental to simuation data, one can note that the general agreement is good, especially if experimental data are averaged by subgrain, which is consistent to what has been done with simulation data. It is also interesting to notice that this difference between grain and subgrain structure is much more significant than for an initial equiaxed microstructure (as shown in previous section \ref{subsec:ModelDRX}).

\begin{figure}[h!]
    \centering
    \begin{subfigure}{0.49\textwidth}
        \centering
        \includegraphics[width=0.95\linewidth]{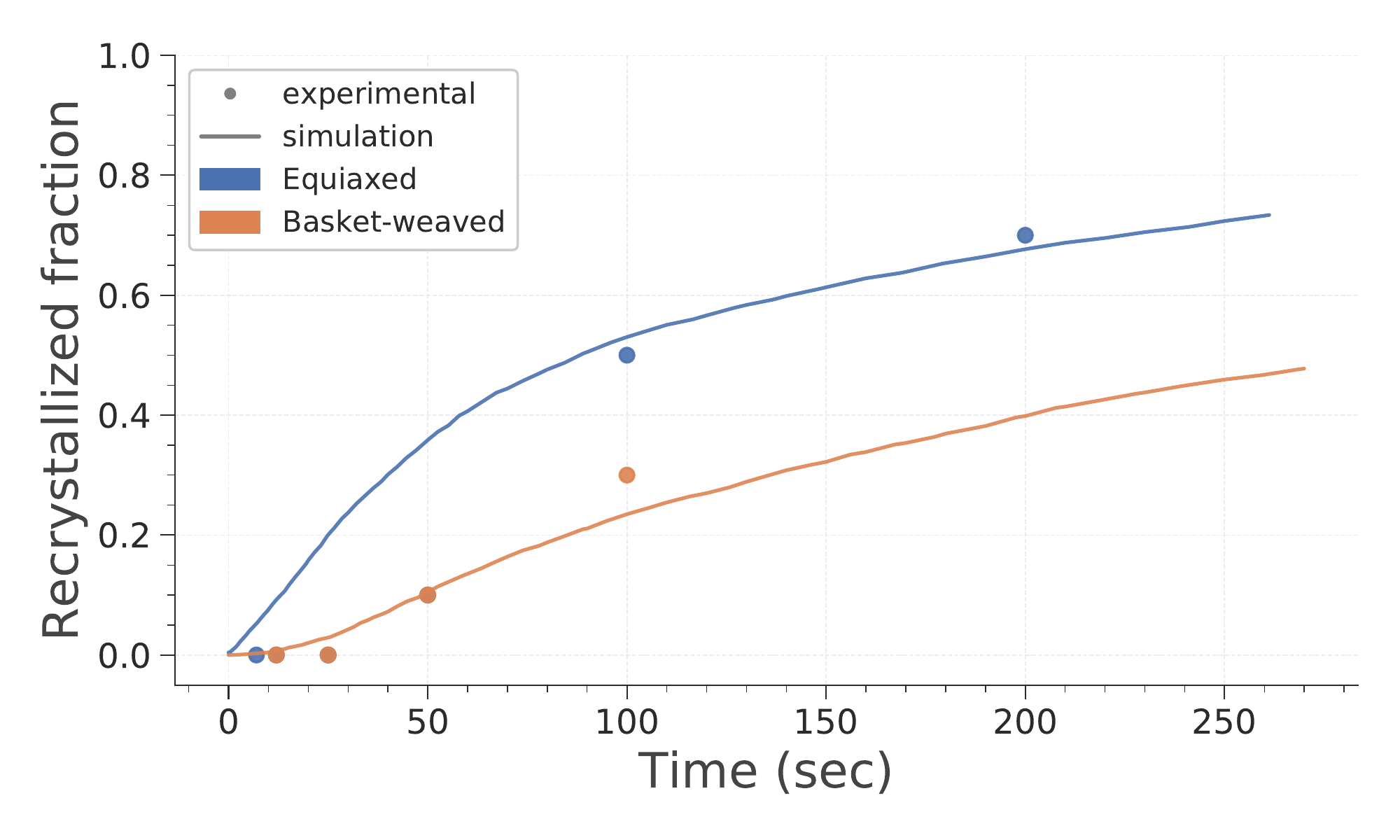}
        \caption{\label{fig:RxFractionFunctionOfTimePDRX} Evolution of recrystallized fraction with time.}
    \end{subfigure}
    \begin{subfigure}{0.49\textwidth}
        \centering
        \includegraphics[width=0.95\linewidth]{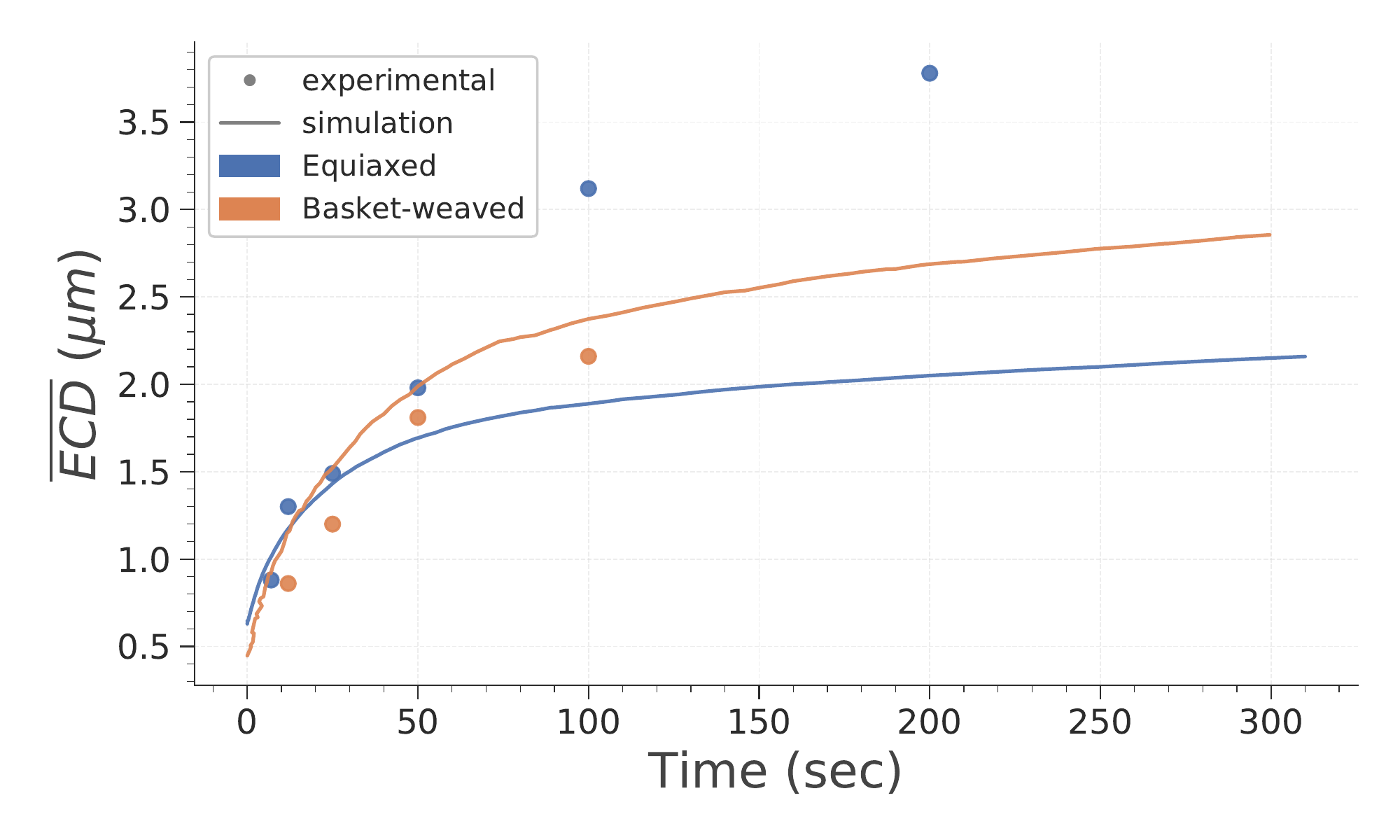}
        \caption{\label{fig:ECDFunctionOfTimePDRX} Evolution of RX grain $\overline{ECD}$ with time.}
    \end{subfigure}
    \begin{subfigure}{0.49\textwidth}
        \centering
        \includegraphics[width=0.95\linewidth]{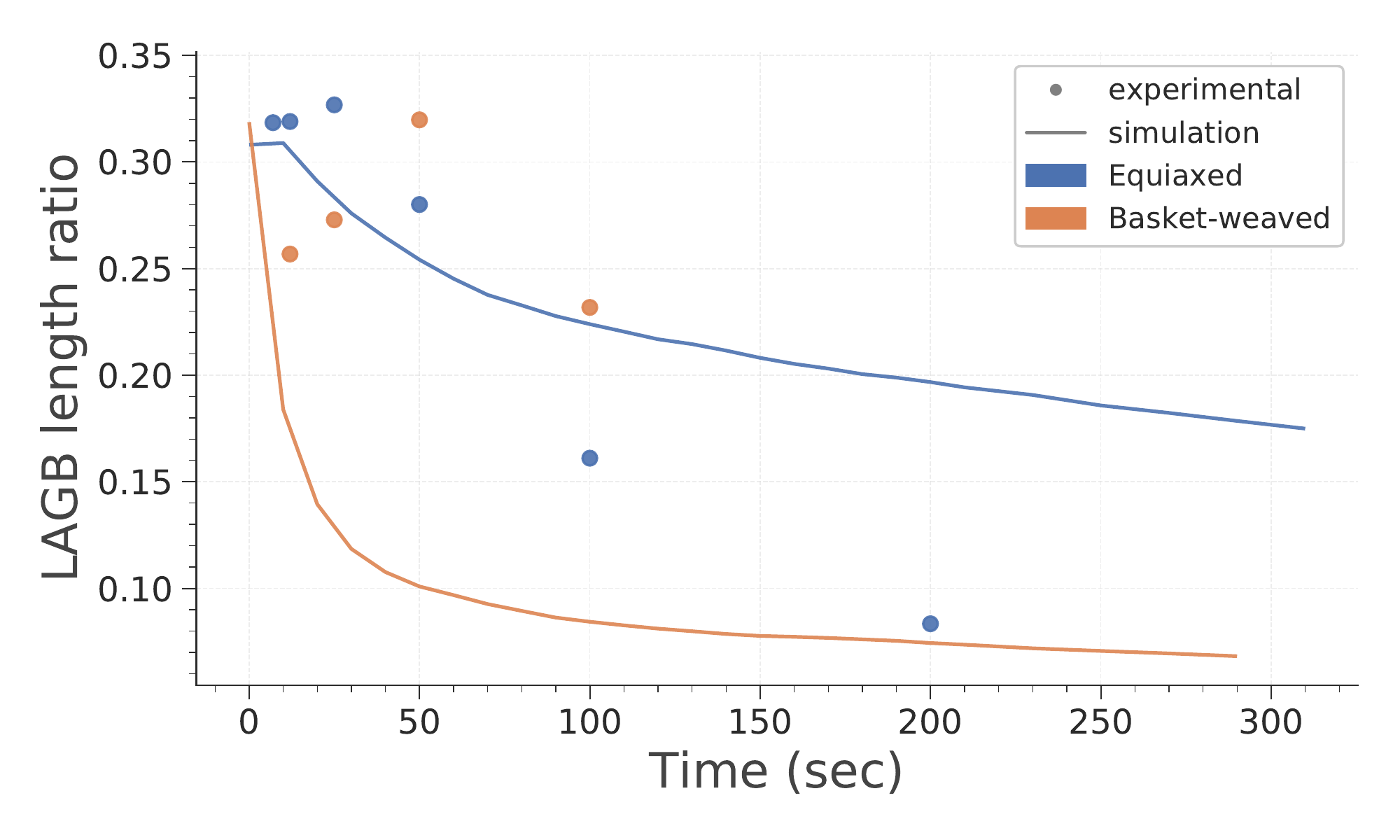}
        \caption{\label{fig:LAGBFractionFunctionOfTimePDRX} Evolution of closed LAGB length fraction with time.}
    \end{subfigure}
        \begin{subfigure}{0.49\textwidth}
        \centering
        \includegraphics[width=0.95\linewidth]{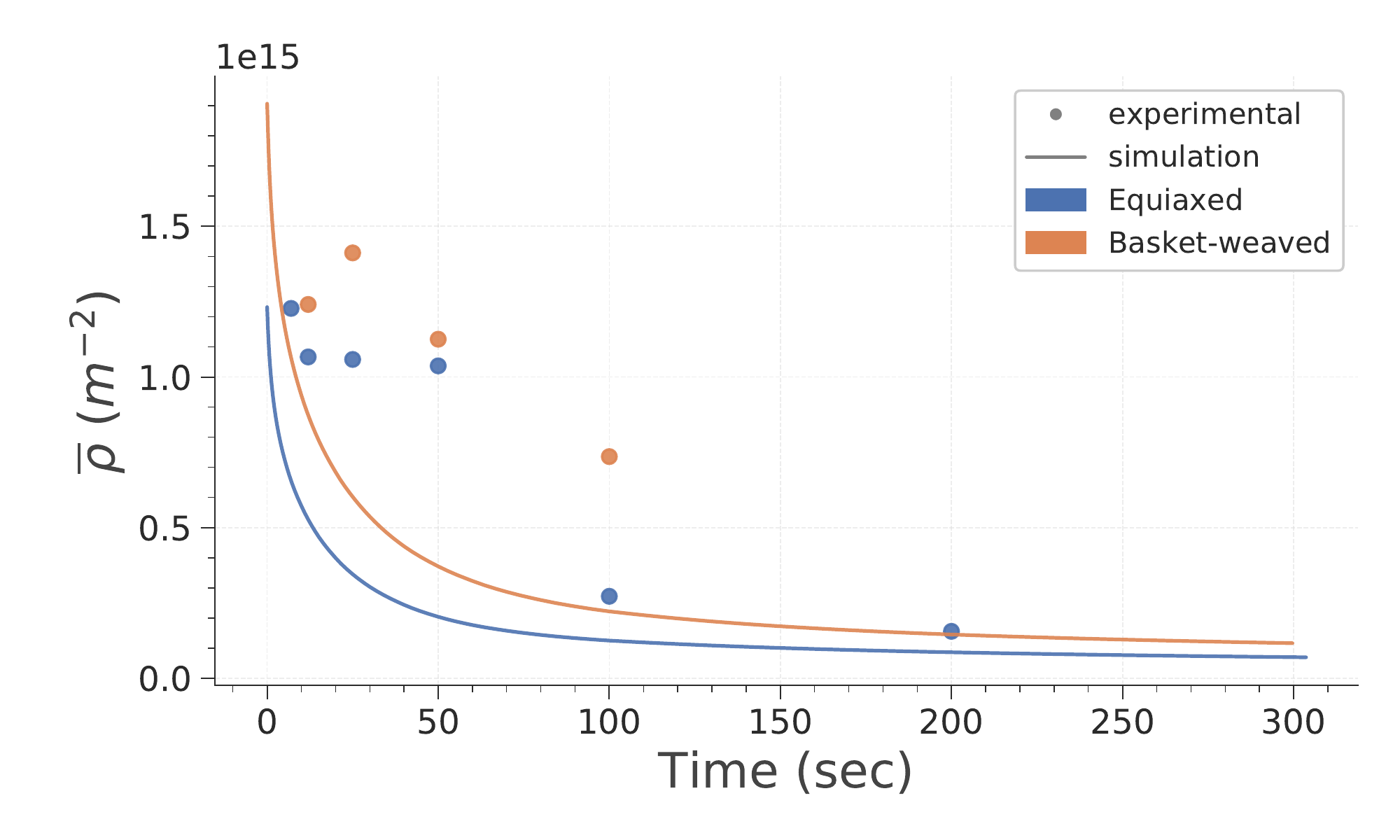}
        \caption{\label{fig:RhoFunctionOfTimePDRX} Evolution of $\overline{\rho}$ with time.}
    \end{subfigure}
\caption{Evolution of the main microstructural features during heat treatment after hot deformation simulation.}\label{fig:NumVsExpGraphsPDRX}
\end{figure}

Figure \ref{fig:NumVsExpGraphsPDRX} presents the evolution of the main microstructure statistical descriptors. One can observe that the numerical results for recrystallized fraction are in good agreement with the experimental ones. It is especially interesting to note that the model is able to reproduce the slower recrystallization of basket-weaved microstructure.  This could appear counter-intuitive looking at figure \ref{fig:RhoFunctionOfTimePDRX} since dislocation density which constitutes the main driving force for recrystallization is higher for basket-weaved microstructure. However, two other phenomena could explain this difference. First, it appears that the LAGB length ratio (see figure \ref{fig:LAGBFractionFunctionOfTimePDRX}) of the basket-weaved microstructure exhibits the same evolution but delayed in time. This could potentially be explained by the difference of spatial distribution of SPP. Indeed, alignment of SPP at lamellae borders in quenched sample could reduce the dislocation mean free path and limit their ability to form LAGB. Another hypothesis is that the increase in deformation localization for quenched microstructures could lead to preferential zones for the formation of recrystallized grains. Those grains would then impinge themselves more quickly and this would lead to a slower recrystallization kinetics.

One should also point out that the model ability to evaluate evolution of other characteristics such as mean dislocation density and LAGB length ratio is good but slightly less precise. Several hypothesis could explain those differences:
\begin{itemize}
\item the model used consider only averaged values by grain. Thus, it is not able to capture the influence of local variations in dislocation density or crystallographic orientation. It also only supports the formation of LAGB which bounds perfectly closed and spherical subgrains which is not seen experimentally. Consequently, this could reduce its ability to predict values related to LAGB.
\item As it is used for producing the results presented here, no static recovery mechanisms are considered. This, as well as the $\rho_0$ value associated to areas swept by moving boundaries could influence significantly driving forces for microstructure evolutions.
\item Finally, no nucleation during the PDRX stage is modeled. There are no clear experimental evidences that this happens for the conditions studied but it remains a probable phenomenon that could influence post-dynamic evolutions. 
\end{itemize}

Except the hypothesis related to intragranular heterogeneities and oversimplified deformation predictions, which are outside the scope of this study, the other will be evaluated and discussed in an upcoming article.

Finally, the consistency of the results obtained either using the CDRX and PDRX framework or only the PDRX one with immersed microstructure is noticeable (see figures \ref{fig:NumVsExpGraphsCDRX} and \ref{fig:NumVsExpGraphsPDRX}). This tends to confirm that the CDRX framework provides precise enough results and that the general simulation environment used is robust and consistent. 

\clearpage
\section{Conclusion}\label{sec:Conclusion}

The present work has detailed some of the main outcomes obtained from an extensive experimental and numerical study of recrystallization of Zy-4. The main recrystallization mechanisms have been identified and described and the influence of some material and thermomechanical parameters quantified. Additionally, the mechanisms observed experimentally have been successfully reproduced by the way of full-field simulations. Adaptation and implementation of the Gourdet-Montheillet model into the LS framework has shown satisfying results. Its ability to reproduce some intergranular heterogeneities and to capture to some extent the evolution of some specific physical quantities related to LAGB has been proven.
Future works will be dedicated to model CDRX with various initial microstructures and for different thermomechanical conditions. A special attention will be paid to obtain an even better agreement with experimental measurements by refining some parameters as discussed in section \ref{subsec:ModelPDRX}. Finally, those numerical results will be analyzed in details in order to assess the influence of some specific mechanisms such as recovery and formation of subgrains in the post-dynamic regime.

\newpage


\bibliography{ASTMproceeding.bib}

\end{document}